# Towards non-threaded Concurrent Constraint Programming for implementing multimedia interaction systems

Mauricio Toro [1]

Universidad Eafit

## Abstract

In this work we explain the implementation of event-driven real-time interpreters for the Concurrent Constraint Programming (CCP) and Non-deterministic Timed Concurrent Constraint (NTCC) formalisms. The CCP interpreter was tested with a program to find, concurrently, paths in a graph and it will be used in the future to find musical sequences in the music improvisation software Omax, developed by the French Acoustics/Music Research Institute (IRCAM). In the other hand, the NTCC interpreter was tested with a music improvisation system based on NTCC (CCFOMI), developed by the AVISPA research group and IRCAM. Additionally, we present GECOL 2, a wrapper for the Generic Constraints Development Environment (GECODE) to Common LISP, developed to port the interpreters to Common LISP in the future. We concluded that using GECODE for the concurrency control avoids the need of having threads and synchronizing them, leading to a simple and efficient implementation of CCP and NTCC. We also noticed that the time units in NTCC interpreter do not represent discrete time units, because when we simulate the NTCC specifications in the interpreter, the time units have different durations. In the future, we propose forcing the duration of each time unit to a fix time, that way we would be able to reason about NTCC time units as we do with discrete time units.

## 1 Introduction

Multimedia Semantic Interaction (MSI) studies systems acting together with the user in order to learn, maintain and update a model of user's behaviour. Examples of this models are: "Style machines" for stylistic motion synthesis [10], a model for music improvisation called OMAX [9] and "A concurrent constraints factor oracle model for music improvisation" (CCFOMI) [44]. Multimedia interaction is studied by many disciplines such as: signal processing, artificial intelligence, computer graphics and concurrency theory. In MSI we are interested in studying the communication of the agents involved in the system, the messages passed between the agents and proving invariants of the system.

Our work is in a broad sense, the exploration of the possibilities and limits of modelling and implementing multimedia interaction systems following the principles

1 email: mtorobe@eafit.edu.co

of a formal model developed by Vijay Saraswat, named Concurrent Constraint Programming (CCP [46]). In CPP, we represent a concurrent system in terms of constraints of the system variables and in terms of agents who reason about partial information provided by those variables. We are convinced that CCP models are appropriate for multimedia interaction, because they have a well defined semantic, allowing us to express synchronization in an easier way than most programming languages and proving properties from the modelled systems.

CCP belongs to a bigger family of formalisms called process calculi. Process calculi has been applied to the modeling of spatially-explicit ecological systems [38, 67, 39, 66] and interactive multimedia systems [4, 65, 61, 34, 59, 56, 58, 60, 7, 64, 57, 62, 63, 55] .

The AVISPA research group have used extensions of CCP to analyse different kinds of interactive systems. For multimedia interaction systems, the reader can look at the models for music improvisation [44], audio processing [13] and formalizing musical processes [45] developed by Assayag, Rueda and Valencia. In other fields, the reader can look at the models from Olarte and Valencia developing an extension of CCP with applications to security [35]; Olarte, Perez, and Rueda modelling biological systems ([21], [33]); and Lopez, Perez, and Rueda verifying properties of security protocols [27]. Further than just modelling those systems, many of them have been successfully simulated using a variety of software tools developed by the French Acoustics/Music Research Institute (IRCAM) and AVISPA, unfortunately none of them are capable of real-time performance.

There are several simulation tools for CCP extensions. Rueda, in collaboration with IRCAM, has developed an interpreter [44] in Common LISP for the Non deterministic Time Concurrent Constraint (NTCC [31]) calculus. Using Mozart-Oz [41], a programming language following the nature of CCP, AVISPA developed a NTCC interpreter called *ntccSim* [20]. Muñoz and Hurtado made an abstract machine in the C language for NTCC, called *Lman* [30]. A few years before AVISPA and IRCAM write their first interpreter for NTCC, Boer, Gabbrielli and Meo used Mozart-Oz to build an interpreter [53] of a Timed Concurrent Constraint (TCC [16] ) language. We give details about the interpreters developed by IRCAM and AVISPA in chapter 2. Furthermore, we present our own implementation of CCP and NTCC capable of real-time performance, inspired by the non-threaded implementation of *Lman* and providing multiple constraint systems and derivate NTCC processes like *ntccSim* and *Rueda's* interpreters.

After considering multiple alternatives to develop those intepreters, we chose to use the Generic Constraint Development Environment (GECODE [47]), written by Christian Schulte, to represent the constraint system and the concurrency control, leading to an *event-driven* (concurrent, but non-threaded) implementation of the CCP and NTCC models. Using GECODE, we avoided the need of using threads. We also explored the possibilites and limitations of developing a generic implementation of *lightweight threads* [28] for Common LISP to write the interpreters.

In chapter 3, we present some alternatives to develop an efficient and generic implementation of *lightweight threads* for Common LISP. Chapter 4 explains the problem of using threads with GECODE and GECOL 2 to program a CCP interpreter. We claim that using GECODE for the concurrency control avoids the need of having threads and synchronizing them, leading to a simple and efficient implementation of CCP and NTCC.

Afterwards, in chapter 4, we also explain how we built two non-threaded interpreters for CCP, one in Common LISP and the other one in C++, writing on top of them a program to find paths in a graph. In Chapter 5, following the design principles of the CCP interpreter, we present *Ntccrt* [61], a generic real-time NTCC interpreter written in C++ and on top of it, an implementation of *CCFOMI*. In order to integrate *Ntccrt* with Common LISP, we extended an existing wrapper for GECODE for Common LISP.

In chapter 6, we explain the architecture the GECOL 2 library, an extension developed to the GECODE wrapper for Common LISP ( GECOL [52]). Finally, to show how can GECOL 2 be used to model musical problems, we present an efficient implementation to find solutions to the *Klumpenhouwer networks* (*k-nets* [26]), modelled as a finite domain Concurrent Satisfaction Problem (CSP) and implemented in both, C++ and Common Lisp, using GECODE and using GECOL 2 respectively.

The reader should be aware that this document is self-contained, meaning that the background explains the basic theory of CCP, NTCC, Factor Oracle, *lightweight threads* and *k-nets*. Additionally, the reader should know that we present future work for each of the chapter separately and finally we present a few conclusions from the whole research.

## 2 Background

## 2.1 Lightweight threads

In computer science, a *continuation* is an abstraction of the processor registers, the *events* are an abstraction of the hardware interruptions and a *thread* represents a sequential flow control or an abstraction of a processor.

Sometimes, *threads* are described by their weight, meaning how much contextual information must be saved for a given thread in order to schedule them [71]. For example, the context of a Unix process includes the hardware register, the kernel stack, user-level stack, process id, and so on. The time required to switch from one Unix process to another is large (thousands of microseconds), for that reason those are called *heavyweight threads*.

Modern operating systems kernels, such as Mac OS X and Mach, allow to have multiple threads in the same process, decreasing the amount of context that must be saved with each one. These *threads* are called *medium weight threads* and it takes hundreds of microseconds to switch between them [54].

When all context and *thread* operations are exposed at user level, each application needs only a minimal amount of context information saved with it, so that context switching can be reduced to tens of microseconds. These are called *lighweight threads*. For example, *lightweight threads* used by the *Java VM* outperform linux *threads* on *thread* activation and synchronization because *thread* management operations do not need to cross kernel protection boundaries. But, linux native *threads* have better performance on I/O operations [6]. Additionally, since *lightweight threads* may block all the other *threads* when performing a blocking I/O operation, it is necessary to use asynchronous I/O operations, adding complexity and increasing the latency for I/O operations.

Strategies to implement *Lightweight threads* include, but are not limited to: Scheduler activations [5], a threading mechanism that maps N user level *threads* into some M kernel *threads*; Protothreads [17], an abstraction that reduces the complexity of *Event-based programs*; virtual machine with thread support [12], supporting the concurrent execution of multiple *threads* in the traditional way; Coroutines [24], allowing multiple entry points, suspending and resuming execution at certain locations; Continuations ([15], [50]), an abstraction of the processor registers commonly used in functional languages; multiple stack based *threads* [69], where we have an scheduler in charge of providing a fair execution to all *threads*; and *Event driven programming* [18], the approach we have chosen to manage concurrency in the interpreters, explained with detail bellow.

*Event-based programs* are typically driven by a loop that polls for events and executes the appropriate call-back when the event occurs. This means that the flow of the program is determined by sensor outputs, user actions or messages from other programs. In order to implement this model it is required: a *dispatcher*, taking the events and calling the appropriate handler; an *event queue*, storing the events when the *dispatcher* is busy; and different *handlers* for each type of events [18] (see figure 1).

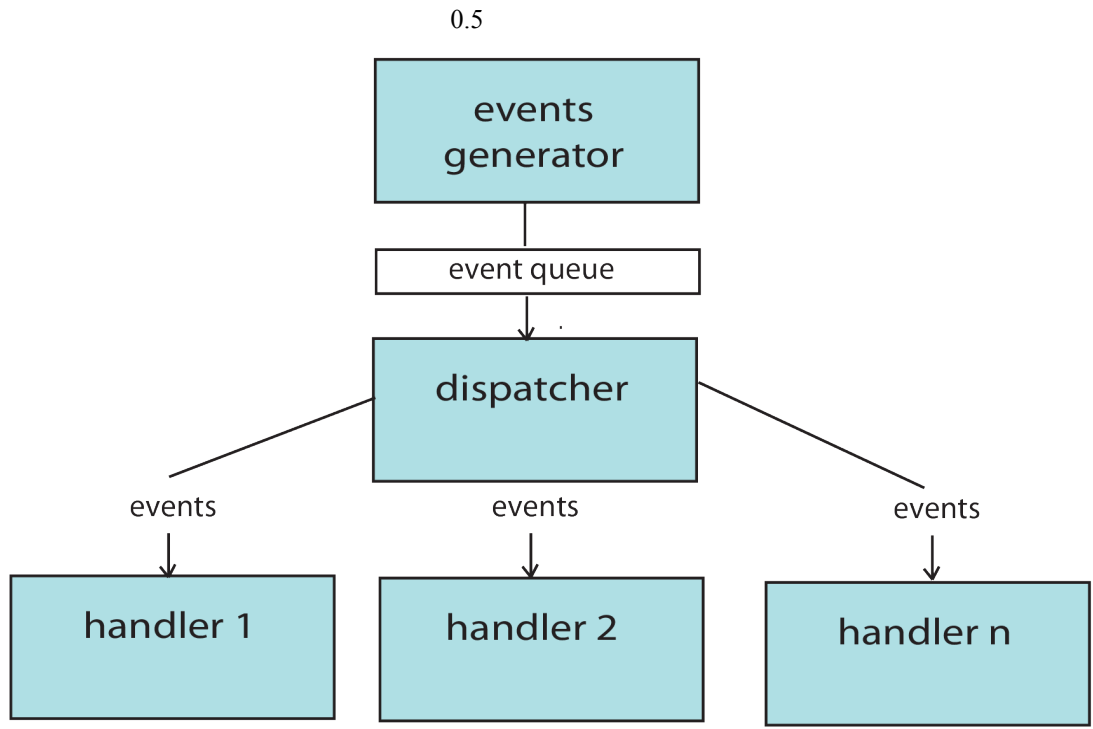

Figure 1: Event Driven Programming Control Flow

## 2.2 Concurrent Constraint Programming

Concurrent Constraint Programming (CCP [46]) is as a model for concurrent systems. In CCP a concurrent system is modelled in terms of constraints over the system variables and in terms of agents interacting with partial information obtained from those variables. This systems provide a *propagator* for each constraint. *Propagators* can be seen as operators reducing the set of possible values for some variables. A constraint is a formula representing partial information about the values of some of the system variables. For example, in a system with variables , taking MIDI values, the constraint specifies possible values for and (where is at least a tone higher than ). The CCP model includes a set of constraints and a entailment relation ⊢ between constraints. This relation gives a way of deducing a constraint from the information supplied by other constraints. For example, (provided in [44]).

"Computation in the CCP model proceeds by accumulating information (i.e. constraints) in a *store*. The information specifies all that is known about the values of the variables at a given moment. Information on the *store* may increase but it cannot *decrease*. Concurrent processes interact with the *store* either *telling* new information or *asking* whether some constraint can be deduced (entailed) from the information contained in it. It may well happens that the constraint cannot be entailed. In this case the interacting process is said to *block* until some other processes tell enough information to the store to deduce its constraint" [44].

Figure 2 presents 4 agents interacting concurrently, the processes and add new information to the *store*. The processes do P and do Q launch process *P* and *Q* respectively, when their condition can be entailed from the *store*. The reader may notice that process do P launches process *P*, but the process do Q will be suspended until its condition can be entailed from the *store* (see figure 3).

## 0.6 Figure 2: Process interaction in CCP

Formally, "a CCP model is based on the idea of a constrain system. A constraint system is a structure <$D$,⊢,*Var*> where D is a (countable) set of primitive constraints (or tokens), ⊢∈$DxD$ is an inference relation (logical entailment) that relates tokens to tokens and *Var* is an infinite set of variables. A (non primitive) constraint is an entailment closed subset of *D*" [46].

"Notice that this definition does not specify particular types of forms of primitive constraints. A constraint systems can thus be adapted to many needs depending on the set $D$ for example expressions such as $x \in R$, where $R$ is a set of ranges of integers (finite domain constrains), or they can be expressions over trees, graphs, sets, etc. In fact, a CCP language usually includes several constraint systems" [43].

## 0.6 Figure 3: Process interaction in CCP (2)

"One drawback of the CCP model as presented above is that information is always accumulated. There is no way to eliminate it. This poses difficulties for modelling reactive systems in which information on a given variable changes depending on the interactions of a system with its environment, as is the case, for example, in interactive musical improvisation systems" [45].

## 2.3 Non-Deterministic Timed Concurrent Constraint Calculus

"This calculus introduces the notion of time, seen as a sequence of time units. At each time unit a CCP computation takes place, starting with an empty store (or one that has been given some information by the environment). Concurrent constraints agents operate on this store as in the usual CCP model to accumulate information into the store. As opposed to the CCP model, however, the agents can schedule processes to be run in future temporal units. In addition, since at the beginning of each time unit a new store is created, information on the value of a variable can change from one unit to the next" [44].

The Non deterministic Time Concurrent Constraint [31] calculus has been used in the past to model multiple kind of systems. For multimedia interaction systems, the reader can look at the models for music improvisation ([44]), audio processing [13] and formalizing musical processes [45] developed by Assayag, Rueda and Valencia. In other fields, the reader can look at the models from Olarte and Valencia developing an extension of CCP with applications to security [35]; Olarte, Perez, and Rueda modelling biological systems ([21], [33]); and Lopez, Perez, and Rueda verifying properties of security protocols [27]. Further than just modelling those systems, many of them have been successfully simulated using a variety of software tools developed by AVISPA and the French Acoustics/Music Research Institute (IRCAM), unfortunately none of them are capable of real-time.

The computational agents of NTCC are described in table 1. We also present some examples about how to use the computational agents for modelling music interaction. Using the **tell** agent it is possible to add constraints such as (meaning the must be equal to 60) or (meaning that is an integer between 60 and 100).

The **when** agent can be used to describe how the system reacts to different events, for example **when do tell**(*CMayor=true*) is a process reacting as soon as the pitch sequence C, E, G (represented as 48, 52, 55 in MIDI notation) has been played, adding the constraint *CMayor=true* to the *store*.

Parallel composition allows us to represent concurrent processes, for example **tell** | **when do tell** (*Instrument*=1) is a process telling the store that is 62 and concurrently reacts when is in the octave -1, assigning the *instrument* variable to 1 (acoustic piano in MIDI notation), as the reader can see in figure 4.

The **next** agent is useful when we want to model variables changing through time [2], for example **when do next tell** , means that if is equal to 60 in the current time unit, it will be different from 60 in the next time unit.

 0.6  Figure 4: tell, when and parallel agents in NTCC

The **unless** agent is useful to model systems reacting when a condition is not satisfied or it cannot be deduced from the store information. For example, **unless next tell** (*lastpitch*<>60), reacts when is different from 60 or it cannot be deduced from the store (i.e. was not played in the current *time unit*), telling the *store* in the next *time unit* that *lastpitch* is not 60. Figure 5 helps to clarify this example.

---

2 This is not possible in CCP because the store is monotonic.

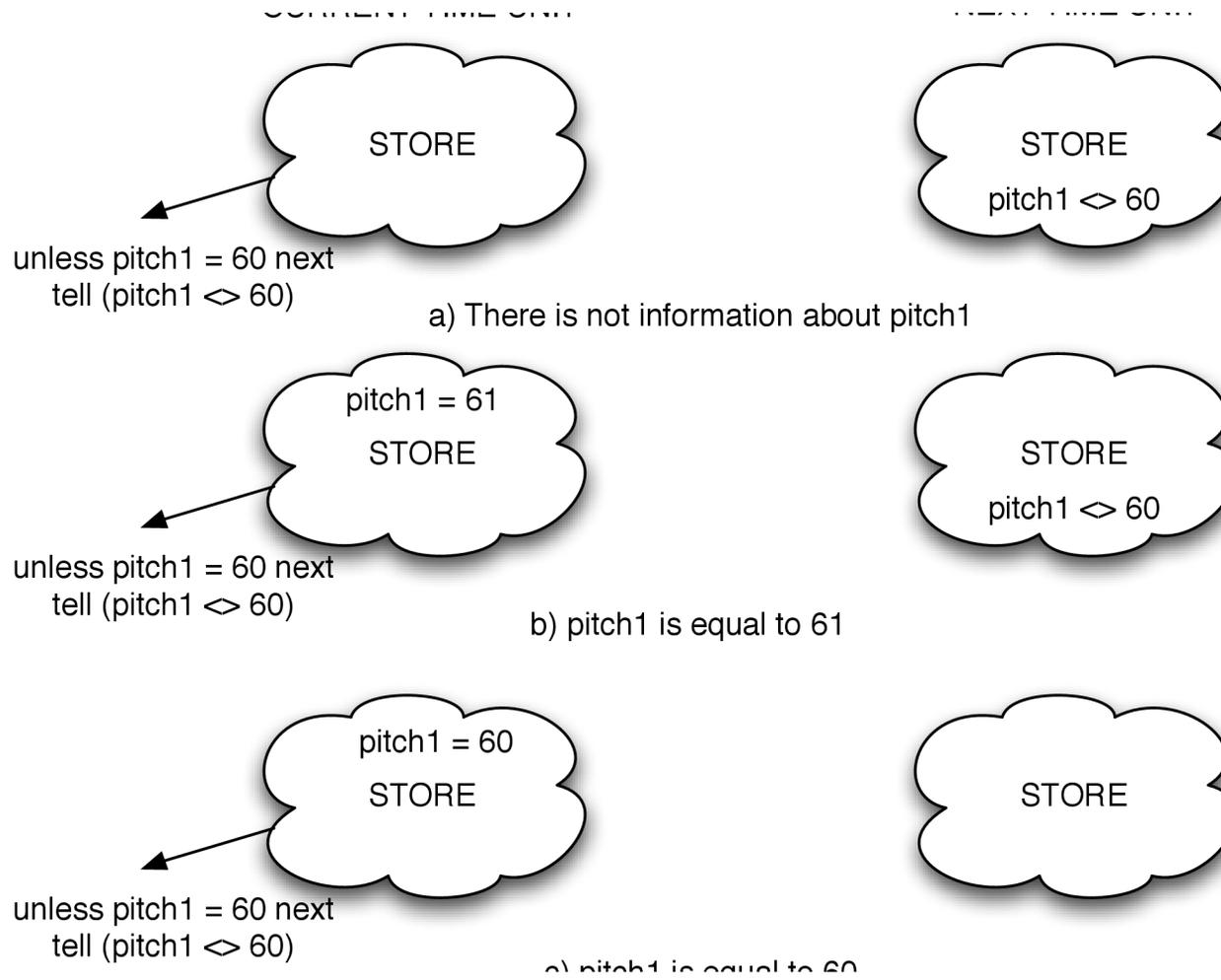

Figure 5: unless agent in NTCC

*P* can be used to delay the end of a music process indefinitely, but not forever *tell (End=true)* and !*P* to execute the process P each *time unit*, for example !**tell** (*PlaySong=true*). The  agent is used to model non-deterministic choices, for example !  **when** *true* **do tell** (*pitch=i*) models a system where each *time unit*, a note is chosen from the C major Chord (C,E or G) to be played.

A Basic recursion definition in NTCC with the form , where *q* is the process name and  is restricted to call *q* at most once and such call must be within the scope of a "next". The reason of using "next" is that we do not want having an infinite recursion within a time unit. Further information about the encoding and the proof principles for recursion can be found in [31].

| Agent | Meaning |
|---|---|
| **tell** (*c*) | Adds the constraint c to the current store |
| **when** (*c*) **do** *A* | if *c* holds now run *A* |
| **local** (*x*) **in** *P* | runs *P* with local variable *x* |
| *A* \| *B* | Parallel composition |
| **next** *A* | Runs *A* at the next time unit |
| **unless** (*c*) **next** *A* | unless *c* can be inferred now, run *A* |
| **when  do** | Non deterministically chooses  s.t. holds |
| *P | Delays P indefinitely (not forever) |
| ! *P* | Executes P each time unit (from now) |

Table 1: NTCC Agents

The agents presented in table 2 are derived from the basic operators, the agent *A* + *B* non-deterministically chooses to execute either *A* or *B*; the process  changes the value of *x* to *t* in the following *time units*. Figure 6 explains the non-deterministic choice in a visual way (Notice that ! **when** *true* **do tell** (*pitch=i*) can be expressed as **tell** (*pitch*=48) + **tell** (*pitch*=52) + **tell** (*pitch*=55) ).

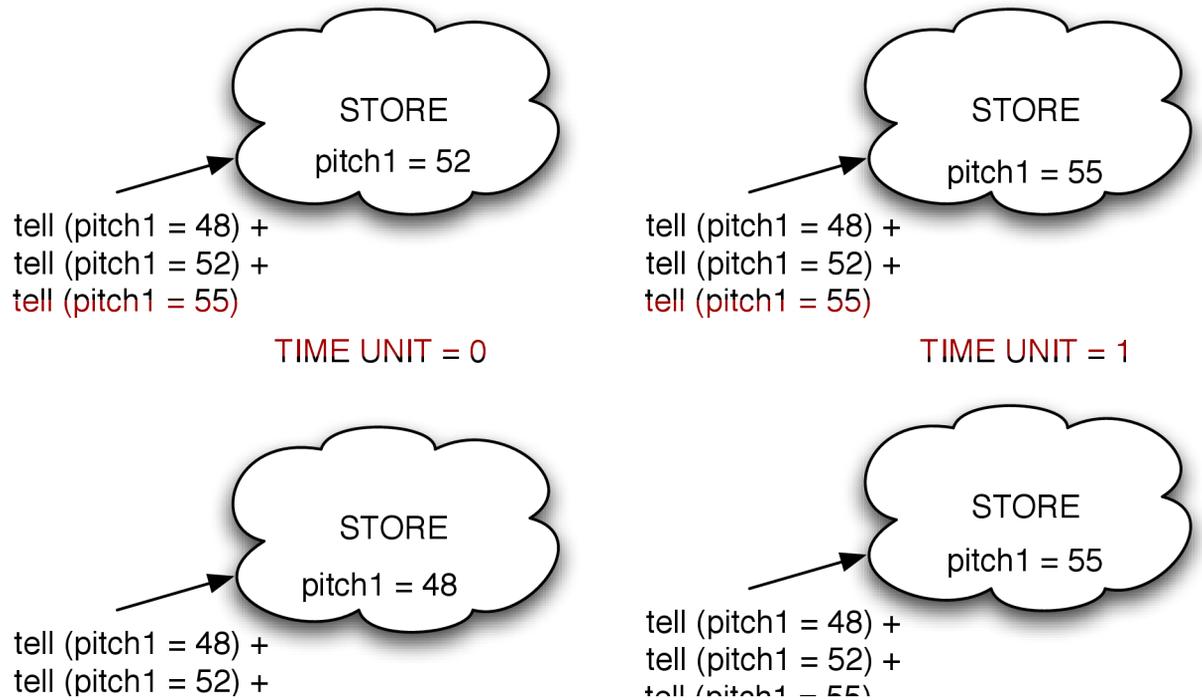

Figure 6: Example of the execution of a non deterministic agent in NTCC

The agents in table 3 are used to model cells, $x:(z)$ creates a new cell $x$ with initial value $z$, changes the value of a cell by the value of a function[3] $g(x)$, such as $g(x)=x+1$ and exchanges the value of cell $x$ and $z$.

| Agent | Meaning |
|---|---|
| A + B | **when** *true* **do** (**when** $i=1$ **do** $A$ \| **when** $i=2$ **do** $B$ ) |
| | **local** $v$ **in when** $t=v$ **do next** ! **tell** $(x=v)$ |

Table 2: Derived NTCC Agents

---

[3] This is different from which changes the value of x only once.

| Agent | Meaning |
|-------|---------|
| *x*: (*z*) | **tell**(*x*=*z*) \| **unless** change(*x*) **next** *x*: (*z*) |
| | **local** *v* **when** *x*=*v* **do** (**tell**(change(*x*) \| **next** (*x*: (*g*(*v*)) ) ) |
| | **local** *v* **when** *t*=*v* **do** (**tell**(change(*x*) \| (**tell**(change(*y*) |
| | \| **next** (*x*: (*g*(*v*)) \| *y*: (*v*)) ) |

Table 3: Definition of cells in NTCC

## 2.4 GECODE

A system providing many efficient propagators and powerful user controllable search engines is GECODE [48]. Written in C++, GECODE includes finite domains (FD) and finite sets (FS) constraint systems. The FD system of Gecode offers many constraint types such as linear constraints ( for example , arithmetic constraints such as *abs*, *min*, *max*, constraints for sorting and for distinct. For FS they provide constraints for relations over sets, set domains, etc.

"State of the art propagators for each constraint type are provided. Each propagator can be seen as a concurrent agent that is inactive most of the time, except when the domain of one of the variables involved in the relation it implements changes. The agent then wakes up and runs its filtering procedure. The whole filtering process involving all propagation agents finishes when a *fix point* (or a failure) is reached. At this point no further modification to the domain of any variable is observed and thus all propagators go to a quiescent state" [43]. A feature of GECODE allowing to developed CCP and NTCC interpreters on top of it, is that the user can ask GECODE to calculate *fix point* at any time calling the *status*() function.

There are also "reified" versions of the above constraints. Given a constraint *c*, a reified constraint asserts , where *b* is a Boolean variable. *Reified constraints* can be used to implement certain kinds of soft constraints. Indeed, when *b* is not bound constraint *c* can be true or false without affecting consistency according to a research report developed by Rueda at IRCAM [43]. The GECODE library includes several search engines (explained in [43]):

- Depth-first search (DFS): Leftmost son is explored first.

- Limited discrepancy search: A value ordering heuristic is assumed. Paths in the search tree are ordered increasingly by the number of node values that are different from the one proposed by the heuristic, and are explored in that order.

- Branch and bound: Solutions are valued (by the user) and computed in increasing order of value.

- Incremental DFS. A DFS that can be restarted from some node. These interact with (possibly user defined) branching procedures that implement different strategies for selecting the next variable and value during search. Most GECODE objects are first class and can be easily refined or extended. In fact, an interface is provided for defining new propagators, search engines or search strategies.

"Another interesting feature is that of computation space. A computation space is a store plus some constraint agents interacting with it. Several computation spaces can be defined for a given problem, either intrinsically by the system (as in the search process) or by the user. It is thus possible to have several different partially solved instances of a given problem. All constraint posting in GECODE refers to some computation space. GECODE is a recently developed system" [43].

## 2.5 Klumpenhouwer networks

Transformational theory is an extension of classic American music set theory, which offers a formalized, mathematical approach to music analysis. The transformational approach, as it is explained by Hascher ([23]), arises from a simple questioning: let *a* and *b* be two musical objects, what do we need to do to *a* in order to obtain *b*? The notions of transformational theory belong principally to group theory, as opposed to "mathematical" set theory on which "musical" set theory is based.

A *Klumpenhouwer network* (*k-net*) is a connected, valued, and directed graph, whose vertices are pitch classes, and whose edges are the operations of transposition and inversion . To explain the intuition of transposition and inversions: let *a,b* be two pitches or elements of the set $\{C,C\#,D,...B\}$. A transposition $a \rightarrow b$ mean that $(a+m)$ *mod* 12=*b*. In the other hand, an inversion means that *b* can be obtained from *a* "reflecting" *a* according to the symmetry line in a *pitch circle* (see figure 8). For example, we can find different *k-nets* for the Pitch class $\{B,F\#,A\}$ as we can see in figure 7.

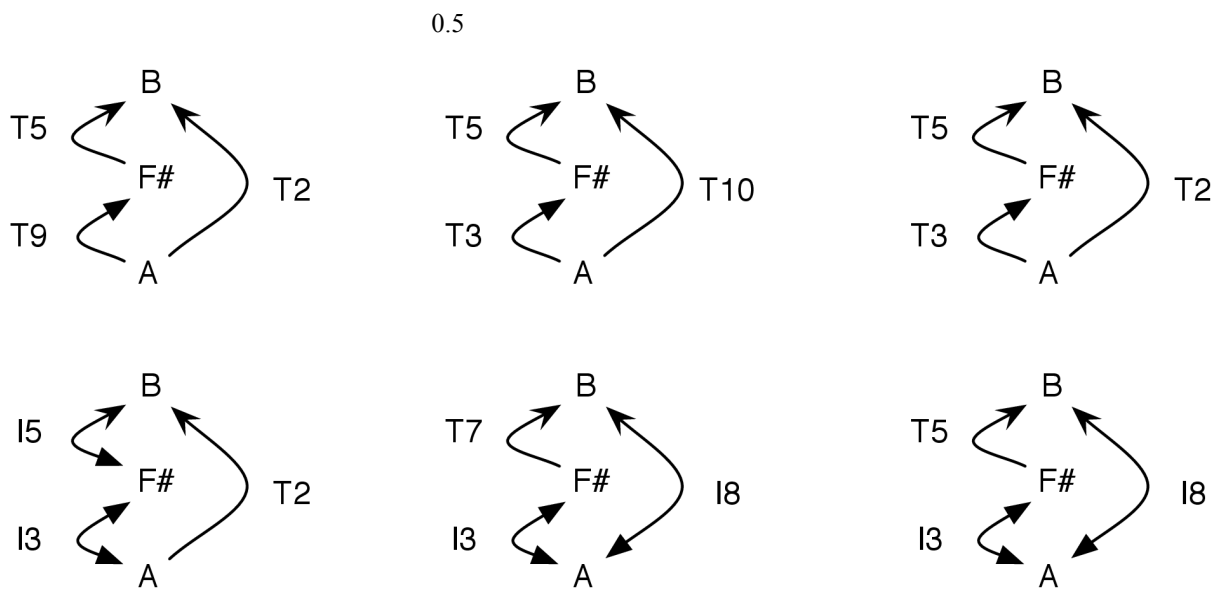

Figure 7: Some K-nets for $\{B,F\#,A\}$

An interesting property of *k-nets* is isography, explained by Hasher as follows: "Let *A,B* be two pitch class sets. *A,B* are in the same canonical classes (or equivalence classes under transposition and inversion) iff it exits *a,b* such that *a* is a *k*-net obtained from *A*; *b* is a *k-net* obtained from *B*; and *a,b* are isographic" (see figure 9). Two *k-nets a* and *b*, displaying the same configuration of vertices and edges, are positively isographic if:

- the disposition of t and i relations is identical in a and b;
- the values *m* of the transpositions tm are the same in a and b;
- the values *n* of the inversions in in b are greater by *k* ($k \geq 1$) than the values *p* of the corresponding inversions in *a*. ([23])

Two *k-nets a* and *b*, displaying the same configuration of vertices and edges, are negatively isographic if:

- the disposition of *t* and *i* relations is identical in *a* and *b*;
- the values *m* of the transpositions are inversely related in *a* and *b*;
- the values *n* of the inversions in *b* are greater by *k* than the inverse of the values p of the corresponding inversions in *a*.

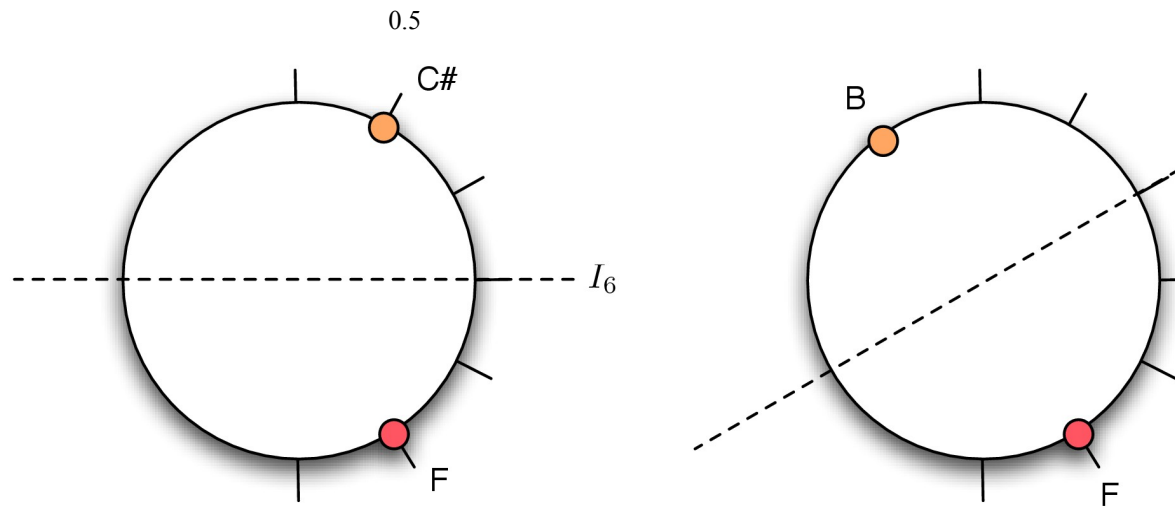

Figure 8: Representing inversions with circles

Isographies can be considered as hyperoperators on *k-nets*, so that for two K-nets *a* and *b*:

- and *b* are positively isographic, with *k=n−p mod* 12
- and *b* are negatively isographic, with *k=n−i(p) mod* 12. ([23])

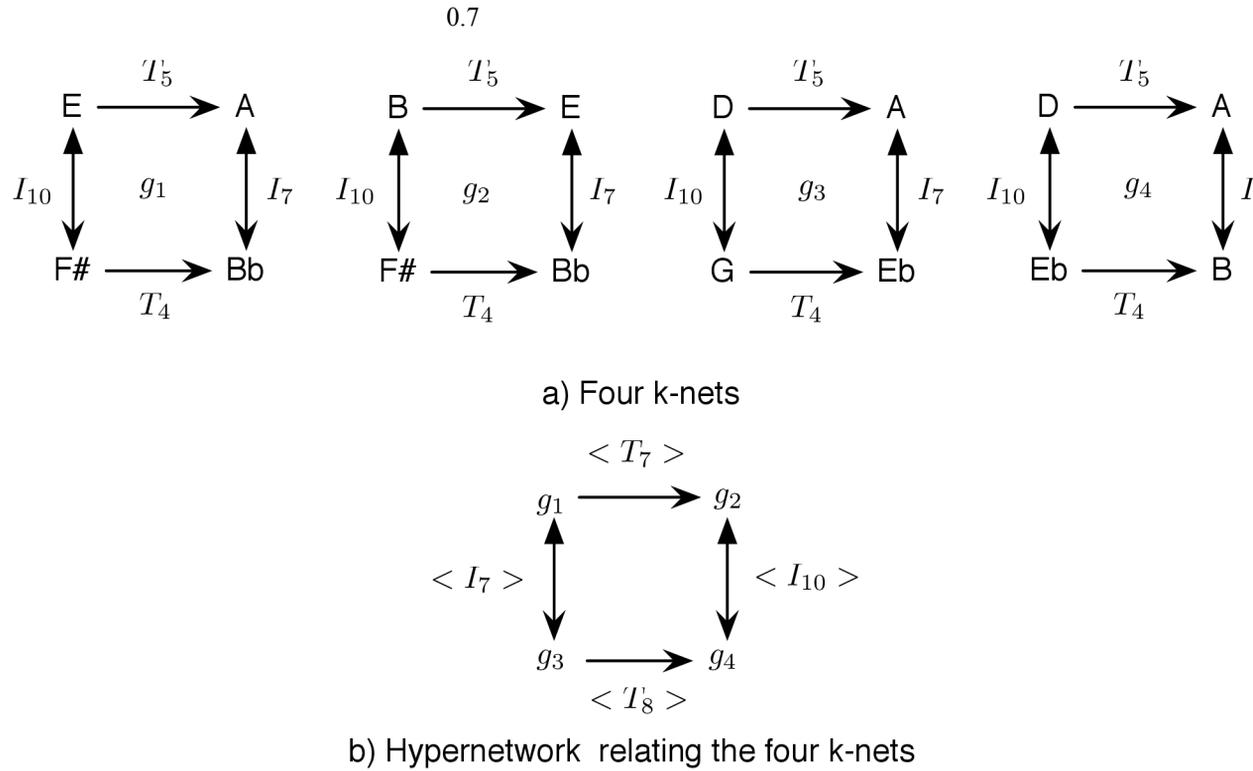

Figure 9: Example of hypernetworks and isographies

## 2.6 Factor oracle

The *factor oracle* (FO) is a finite automaton constructed in linear time and space in an incremental fashion. A sequence of symbols s = is learned in such automaton, which states are 0,1,2...n. There is always a transition arrow (called *factor link*) from the state i - 1 to the state i and there are some directed "backwards", going from state i to j, called *suffix links*, and bear no label. For example, a FO automaton for s = abb is presented in Figure 10, where black headed arrows represent the *factor links* and while headed arrows represent the *suffix links* according to [44]. The formal definitions can be found in [3].

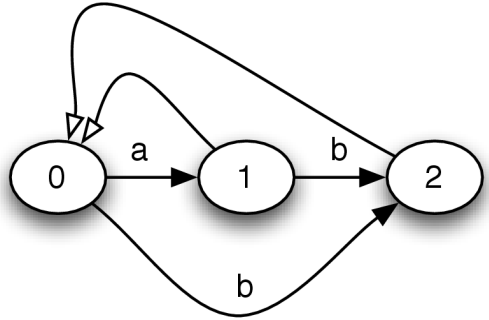

0.6

Figure 10: A FO automaton for s = ab

The oracle is learned on-line and it was proved in [3] that is O(*m*) in time and space. For each new entering symbol σ, a new state *i* is added and an arrow from *i*−1 to *i* is created with label . Starting from *i*−1, the suffix links are iteratively followed backward, until a state is reached where a factor link with label  going to some state j, or until there is no more *suffix links* to follow. For each state met during this iteration, a new *factor link* labelled by  is added from this state to *i*. Finally, a *suffix link* is added from *i* to state *j* or to state 0 depending on which condition terminated the iteration.

FO's were initially conceived for optimal string matching, and were extended easily for computing repeated factor in a word and for data compression [25]. They have been used for music improvisation in *CCFOMI* [44] and Omax [8].

## 2.7  Concurrent Constraint Factor Oracle Model for Music Improvisation

Musical improvisation is the spontaneous creative process of making music while it is being performed. To use a linguistic analogy, improvisation is like speaking or having a conversation as opposed to reciting a written text.

Machine improvisation and related style simulation problems usually consider building representations of time-based media data, such as music, either by explicit coding of rules or applying machine learning methods. We call *Stylistic learning* the process of applying such methods to musical sequences in order to capture salient musical features and organize these features into a model. The *Stylistic simulation* process produces musical sequences stylistically consistent with the learned material [44].

*CCFOMI*, a system proposed in [44], is divided in three subsystems: learning (ADD), improvisation (CHOICE) and playing (PLAYER) running concurrently. In addition, there is a synchronization process (SYNC) and a loop process (LOOP).

The system uses three kind of variables to represent the partially built Factor Oracle automaton: Variables  are the set of labels of all currently existing *factor links* going forward from *k*. Variables  are *suffix* (i.e. backward) *links* from each state i and variable  give the state reached from *k* by following a *factor link* labeled .

Process  adds (if needed) *factor links* labeled  to state *i* from all states *k* reached from *i*−1 by backwards links, then computes , the *suffix link* from *i*.

**when** *k*≥0 **do**
**unless**
**next**(! **tell** () | ! **tell**() | )
| **when** *k*=−1 **do** ! **tell**()
| **when**  **do** ! **tell**()

The Process is the one in charge of adding a new symbols to the automata.

! **tell**() | ()

The two processes above model the learning phase. The learning and the simulation phase must work concurrently. In order to achieve that, it is required that the simulation phase only takes place once the subgraph is completely built. The process is in charge of doing the synchronization between the simulation and the learning phase to preserve that property. This is greatly simplified by the used of constrains. When a variable has no value, *when* processes depending on it are blocked. Therefore, the process is "waiting" until *go* is greater or equal than one which means that the process has played the note *i* and the process can add a new symbol to the FO. The other condition is because the first *suffix link* of the FO is equal -1 and it cannot be followed in the simulation phase.

**when**  **do** ( | **next** )
**unless**  **next**

"A musician is modelled as a process playing some note *p* every once in a while. The process non deterministically chooses between playing a note now or postponing the decision to the next time unit" [44].

**when** true **do** (! **tell**() | **tell**(*go*=*j*) | **next** )
+ (**tell** (*go*=*j*−1) | **next** )

"The improvisation process uses the distribution function Φ : . The process starts from state *k* and stochastically, chooses according to probability *q*, whether to output the symbol or to follow a backward link " [44].

**when**  **do** **next**( **tell** () | )
| **tell** ()
| **when**  **do** **next** (**tell** () | )
| **unless**
**next** **when**  **do** ( **tell** (*out*=σ) |

"The whole system is represented by a process doing all the initializations and launching the processes when corresponding. Improvisation starts after *n* symbols have been scheduled by the player" [44].

! **tell**(*q=p*) | ! **tell**() | |
| ! **when** *go=n* **do** *CHOICE*(*n*)

## 2.8 NTCC Interpreters

### 2.8.1 Lman

This interpreter was developed by the AVISPA research group, in 2003, as a framework to program RCX Lego Robots. It was the first approach to simulate the NTCC calculus and it is composed of three parts: an Abstract machine [30], a compiler [36] and a visual language [19]. It uses the finite domain constraint system [70] used in the Cordial language [40], based in the model of Bjorn Carlson [14]. The architecture used can be seen in figure 11.

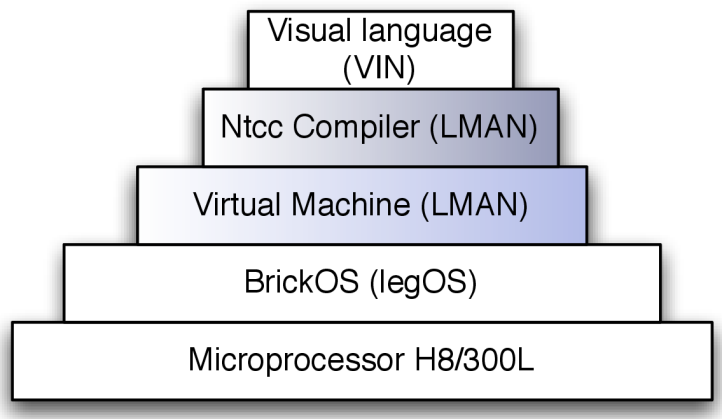

0.6

Figure 11: Architecture used by Lman, the compiler and the visual language working together

This remarkable implementation [30], written in C, was presented in CP2004. This interpreter introduced the idea of having several queues for storing NTCC's processes: *running* queue, *unless* queue and *suspended by ask* queue. Additionally, it stores the variables associated to each process, to know when the processes are not longer blocked. Lman represents constraints as a first order formula using trees. For instance, the constraint ($y$+2)=($x$*4)−($z$/6) can be represented with this binary tree in figure 12.

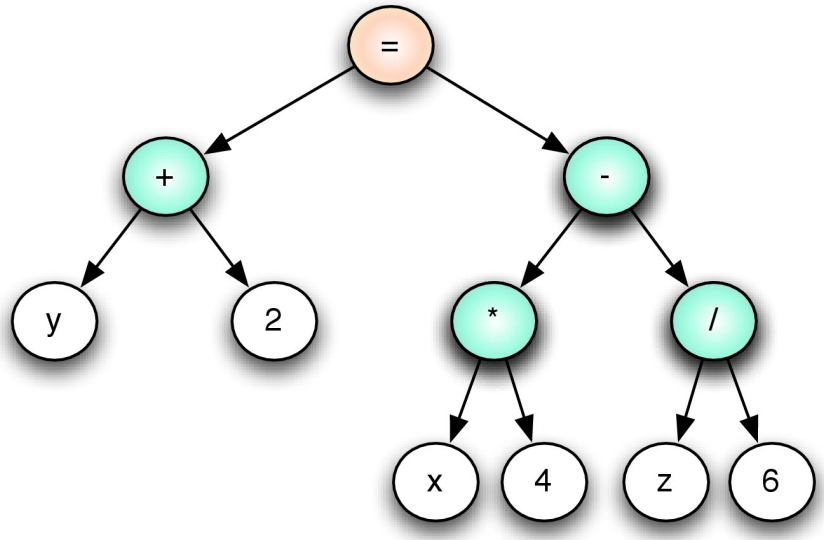

Figure 12: Representing a constraint in Lman

Regrettably, since *Lman* is implemented in the C language, it does not offer abstractions such as objects making very difficult its extension; it only supports finite domain constraints; it was not designed for real time; and finally, the representation of cells is not efficient.

```
// cell's implementation
  (when (ca=1) do ((next tell (ca=1))
  || (unless (changea=1) next ((when (a=0) do (next tell (a=0)))
     + (when (a=FWD) do (next tell (a=FWD)))
     + (when (a=RGHT) do (next (tell (a=RGHT))))
     + (when (a=LFT) do (next ( tell (a=LFT))))))))
```

The representation of cells used by *Lman* is inefficient because it follows the specification of the formalism as we saw it in previous sections. This representation was improved in *Rueda's interpreter* using *Screamer* variables. Those disadvantages led to a correct implementation of NTCC, but unfortunately, not capable of real-time according to the tests presented by the developers.

Following, we present a test made by *Lman*'s developers [29]. The test was made using a Pentium III 930 Mhz, 256 MB Ram, Linux Debian Woody (3.0) and the RCX 2.0 Lego robot with running BrickOS 2.6.1. This application plays a sequence of midi pitches with a fixed durations.

```
   (! tell (osnd1=ON))
|| (! tell(osnddur=HALFNOTE)) --Process 2: Duration of each note
|| (next 1 tell (ovalsnd1=GC)) --Process 3: Playing the note
|| (next 2 tell (ovalsnd1=GE))
...
|| (next 46 tell (ovalsnd1=CC)))
```

This simple process takes an average of 281.25 ms to run each *time unit* using *Lman*, unfortunately it is not suitable for real-time interaction in music, even if we would run it on modern computers.

### 2.8.2 NtccSim

This interpreter was developed by the AVISPA research group, in 2006, as a simulation tool to run NTCC specifications and it was used to simulate biological models [22]. It was developed in Mozart-OZ, providing an easy way to represent recursive procedures. Furthermore, it is able to work with finite domains (FD) and real intervals (XRI) as constraint systems.

There are a few disadvantages of *ntccSim*: it does not provide an abstraction to work with constraint systems such as rational trees and sets used to model the *Factor Oracle* in music improvisation models. Another disadvantage is that even tough, using Mozart-Oz makes the programming of simulation tools for concurrent constraint programming calculi very easy, we conjecture (it has not been proved) that using Mozart-OZ for writing a NTCC interpreter it is not as efficient as using the C++ library GECODE, based on the results obtained in the benchmark examples of GECODE [49] and some results we got with a CCP program tested on Mozart, Lisp and C++ (see chapter 4).

The system shown bellow, calls a recursive definition which acts like a counter, updating the value of each time unit. Additionally, it has a *when* process and an *unless* process running concurrently.

**tell**
| **next**

*System*
**when do tell**
| **unless do tell**
|

This NTCC system can be easily represented in *ntccSim* as follows

```
P3 = when(proc{$ Root} Root.current.x1 >: 3 end
     tell(proc{$ Root} Root.current.x3 =: 6 end))

P5 = unless(proc{$ Root} Root.current.x1 =: 5 end
       tell(proc{$ Root} Root.current.x5 =: 1 end))

P10 = fun lazy{$ X}
  par(tell(proc{$ Root} Root.current.x10 =: X end) next({P10 X+1})) end

{NTCC.simulate [P3 P5{P10 1}]}
```

### 2.8.3 Rueda's interpreter

This interpreter was developed by the AVISPA and IRCAM, in 2006, as a framework to program multimedia semantic interaction applications. This interpreter was the first one representing the rational trees, finite domain (FD) and finite domain sets (FS) constraint systems.

In the interpreter each NTCC agent is represented by MCL processes (efficient *medium weight threads*) to manage concurrency. Each process has a particular waiting function. For example, the waiting function of the *when* processes wait until its guard can be deduced to be true. The *unless* processes wait until the current time unit is done. The waiting function is used by the Lisp scheduler to decide whether to activate or defer the process (the architecture of this interpreter is show in figure 13). Additionally, there is a concurrent *TICK* process permanently testing *stability* of the current time unit. All the mentioned processes run concurrently in a separate thread. Rueda argues that this architecture follows closely the concurrent nature of NTCC [44].

One drawback of this interpreter is the use of *Screamer* [51] (a framework for constraint logic programming written in Common LISP) to represent the constraint systems, making the execution of the NTCC specifications not suitable for real-time interaction, since *Screamer* is not designed to run in real-time (the performance of this system is compared with *Ntccrt* in chapter 5).

The syntax of NTCC processes in the interpreter closely resembles the corresponding calculus definitions (using Lisp parenthesized prefix notation). For example, process  is implemented as

```
(define-process synci (i)
  (let ( (S_i (make-variable "S" i)))
    (parp (whenp (and (>= S_i-1 -1) (>= go i))
                (parp (callp addi i) (nextp (callp synci (1+ i))) )
          (unlessp (and (>= S_i-1 -1) (>= go i) ) (callp synci i)))))
```

where *parp* stands for the NTCC *parallel* construct. Recursive NTCC process definitions are thus simply implemented as lisp functions invoked using a special *callp* primitive. The recursive call only takes effect when the is preceded by a *next* process. [44]

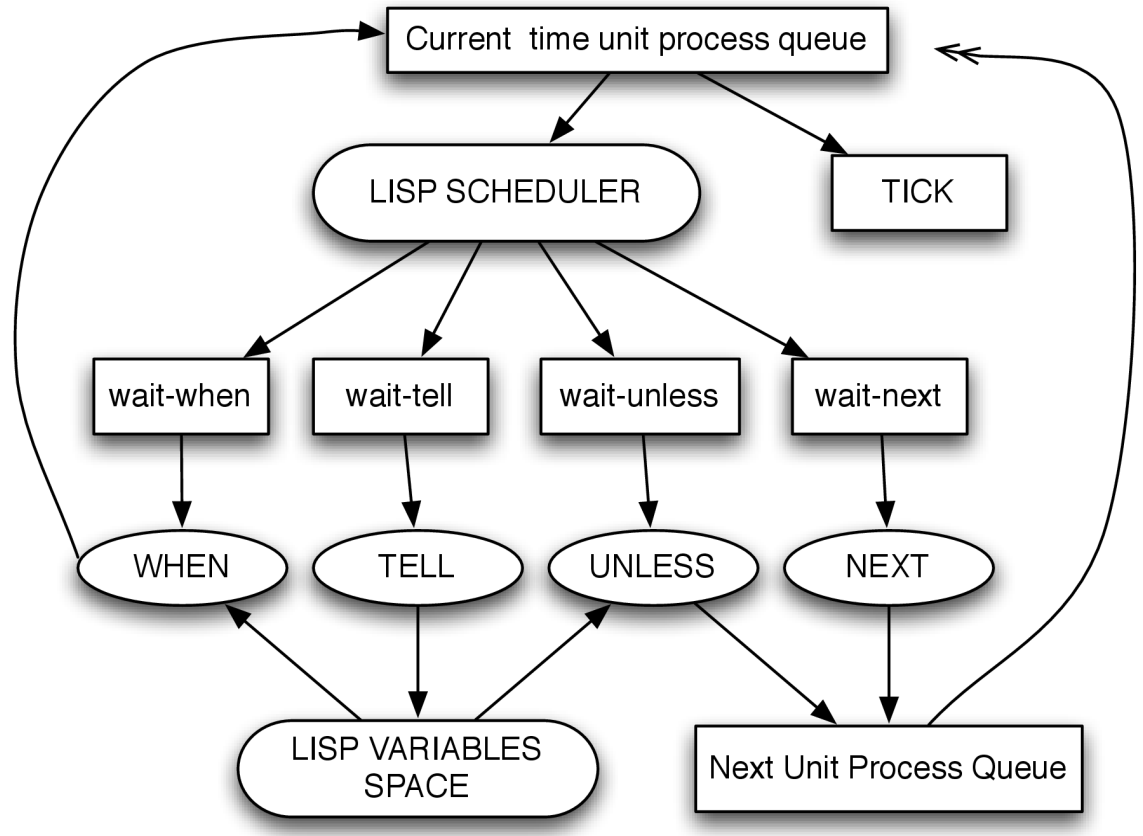

Figure 13: Architecture used by Rueda's interpreter

# 3 Implementing lightweight threads for Common Lisp

*Try Lispworks' simple-processes instead of writing your own threading library. -*
*Gérard Assayag*

OpenMusic [11] (successfully used for constraint programming [1]) and Omax [8] are applications developed by IRCAM in Common LISP. Currently, they use *Lispworks* (a non-free Common LISP distribution). In order to keep our intepreters compatible with OpenMusic and Omax, we have explored the possibility of implementing generic *lightweight threads* for Common LISP, testing their performance in *Lispworks Professional 5.02* under Mac OS X for Intel.

## 3.1 Using continuations to represent threads

> *It is impossible to implement multiprocessing within a portable Common Lisp program, for use by Common Lisp code, except by rewriting code into continuation-*
>
> *passing style. - Kevin Reid*

It is possible to simulate concurrent *threads* using *continuations* and UNIX signals (to provide preemption). The *continuation* of each *thread* must be saved. This way they can be invoked at a later time. When a *thread* must block, we can capture the *continuation* using the *call/cc* macro, provided by libraries such as *cl-cont*, since continuations are not natively implemented in Common LISP. This approach has been used before to implement a concurrent version of *ML* called *SML/NJ* [15].

Using *continuations* posses a few problems. They only capture the state that describes the processor, but they do not capture the state of the I/O systems [50]. Another issue is the lack of a native implementation in Common LISP. Even though, the *continuation passing style (CPS)* can be obtained writing LISP macros, it creates an overhead leading to a high memory and time consumption.

Table 4 describes the memory and time consumption of a function adding all the elements of a list containing 20000 elements. We executed several times both the normal code and the *CPS* code (generated by *cl-cont*) in *Lispworks* and SBCL. Results were obtained after several tests under *Mac OS 10.5* using an Imac Intel Core 2 duo 2.8 ghz, *Lispworks Professional 5.02* and *SBCL 1.012*. We concluded that Lispworks performance is not very good for *CPS* code, probably because the compiler is not optimized to handle *CPS* code. Therefore, we do not recommend using this approach to implement *lightweight threads* for Common LISP.

| Normal code | Lispworks | SBCL |
|---|---|---|
| Time consumption | 0.004 | 0.002 |
| Memory consumption | 5492 | 320466 |
| **CPS code** | | |
| Time consumption | 0.242 | 0.018 |
| Memory consumption | 13600736 | 5763064 |

Table 4: CPS code Vs Lisp code (time in seconds, memory in bytes)

## 3.2 Transforming Common LISP code to Event Driven Programming

> *If you have to write it (lightweight threading library) completely in Lisp with no support at all from the language, then if you want efficiency you will have to change how you write programs. You will have to write each program as an event loop. Peter Van Roy*

Each program will be written as an event loop, which runs by taking an event and executing some code, depending of the type of event, and then posting one or more new events. *Lightweight threads* can be done this way, having several event queues. The scheduler picks one event from one queue to execute each time around [42].

In order to achieve *thread* synchronization, we have a *wait* and a *bind* events working on top of dataflow variables. Asynchronous send and blocking receive can be achieved assigning a mailbox to each *thread* [4].

Our implementation of *lightweight threads* for Common LISP is composed by: a *runnable thread* queue, *current thread* variable, and a hash table to keep a relation between a lock and the *threads* waiting for that lock. The *threads* are modelled by a structure containing an identifier, an status (suspended, running, terminated), a reference to a synchronization variable (when it is suspended), an event queue and a priority.

```
(defstruct thread name status whoamiwaitingfor EventQueue Priority)
```

We also provide 6 simple kind of events: *execute*, *bind*, *wait*, *let*, *waitforlock* and *dotimes* with a handler associated to each of them. The handler for the *execute* event is simple, it evals the instruction encapsulated in this event. Notice that this leaves the responsibility of using it only for simple instructions to the programmer. For example, encapsulating an infinite loop or an infinite recursion inside this event leads to an unfair scheduling. The *bind* and the *lock* events are used for synchronization and the *let* and *dotimes* events help fragmenting blocks containing multiple instructions.

The code bellow represents an implementation of a multithreaded matrix multiplication algorithm, for each multiplication necessary, a new thread is created ( threads are created for two square matrices of size *n*) and synchronization is provided by locks. This is the implementation using *Lispworks* processes as the threading library

```
(defun runrunrun ()
(let ((thelock (mp:make-lock)) (thelock2 (mp:make-lock)))
(dotimes (i n ) (dotimes (j n ) (dotimes (k n )
  (mp:process-run-function "Cik = Aij*Bjk" nil
      (lambda (II JJ KK) (mp:process-lock thelock)
(setf (aref *C* II KK) (+ (aref *C* II KK)  (* (aref *A* II JJ) (aref
*B* JJ KK) )))
                        (mp:process-unlock thelock)  (mp:process-lock
thelock2)
   (setf *counter* (- *counter* 1)) (mp:process-unlock thelock2)) i j
k))))))
```

Contrasting to the implementation above, our implementation uses the *event driven programming* interface described previously. Each *thread* created to make a multiplication does not use locks, since each *setf* instruction is made atomically with the *execute* event. Instead, it uses dataflow *variables* (having two states, bind or not bind) to be synchronized with another thread in charge of knowing when the execution of all the threads is done.

```
(dotimes (i *n*) (dotimes (j *n*) (dotimes (k *n*)
(thread-run-function
 (lambda (i j k)  (setf (thread-EventQueue *current-thread*)  (list
   (make-execute :body (list 'setf (list 'aref '*C* i k)
```

4 Lispworks have an API for mailboxes already implemented

```
    (list '+ (list 'aref '*C* i k) (list '* (list 'aref '*A* i j) (list
'aref '*B* j k)))))
(make-bind :who (aref *sync* i j k))))) (list i j k) 1 'calcul))))
```

This is the thread in charge of knowing when the execution of all threads is done

```
(thread-run-function
(lambda ()
 (let ( (i (newletvariable)) (j (newletvariable)) (k (newletvariable)))
  (setf (thread-EventQueue *current-thread*) (list
   (make-dotimmes :varname i :init 0 :end *n* :step 1 :body
    (make-dotimmes :varname j :init 0 :end *n* :step 1 :body
     (make-dotimmes :varname k :init 0 :end *n* :step 1 :body
      (make-wait :who `(aref *sync* (gethash ,i *letvariables*)
 (gethash ,j *letvariables*) (gethash ,k *letvariables*)))))))))))) nil 1
'wait )
```

Figure 14 compares the execution times of the *Lispwork processes* (native *medium weight threads* provided by *Lispworks*), our implementation of *event driven programming* and the sequential version of the matrix multiplication algorythm in an Intel 2.8 GHz using Mac OS 10.5.2, running *Lispworks* 5.02 *professional*. Additionally, we tested *simple-processes* provided by *Lispworks* multiprocessing API, but they were very slow, taking around 10 seconds for 16 threads. Furthermore, they are very unstable in *Lispworks* 5.0 under Mac OS X, often crashing the whole IDE.

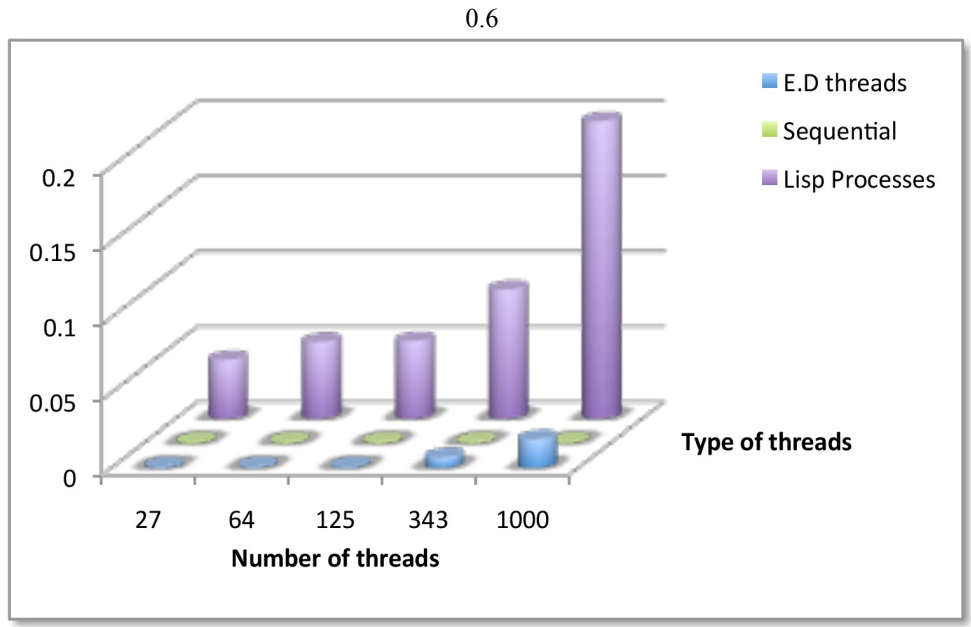

Figure 14: Multithreaded matrix multiplication (time in seconds)

Although transforming Common LISP code to *event driven programming* came up being efficient, the lack of generality of this approach, makes it inappropriate for many applications. For instance, it will be necessary to create events for *go to* jumps, for exception handling, asynchronous signals, loop macros,etc.

## 3.3 Future Work

In the future, in order to use *lightweight threads* in Common LISP, we recommend using an implementation with *lightweight threads* such as *CMU-CL* (http://www.cons.org/cmucl/). Notice that current version of *CMU-CL* (CMU-CL 19e) provides binaries for Mac OS X PPC and Intel.

## 4 Developing CCP interpreters using GECODE

*Ask* processes can be easily represented in GECODE taking advantage of the *reified propagators* and *tell* processes can be represented with *non-reified propagators*. Additionally, it is important to mention that GECODE is not thread safe, being necessary to add locks for all the concurrent reading and writing operations, adding an overhead when using threads. Another fact is the event driven nature of GECODE itself [48], allowing us to express CPP and NTCC in an event driven style, without writing code for a dispatcher nor event queues. In this chapter we will explain the different approaches explored to develop a generic real-time interpreter for CCP.

## 4.1 Applications: Finding paths in a graph concurrently

An application where we use the CCP interpreter is finding, concurrently, paths in a graph. The idea is having one CCP process for each edge. Each sends *forward* "signals" to its successors and *back* "signals" to its predecessors. When an receives a *back* "signal" and a *forward* "signal", it tells the *store* that there is a path and adds *j* to the set . After propagation finishes, we iterate over the resulting sets to find different paths. For instance, we can build a path in the graph getting the lower bound of each set.

**when do (tell () | tell () )**
| **when do tell ()**
| **when do tell ()**

The *Main* process finds a path between the vertices *a* and *b* in a graph represented by *edges* (a set of pairs (*i,j*) representing the graph edges). The *Main* process calls for each (*i,j*)∈*edges* and concurrently, it sends *forward* "signals" to processes with the form  and *back* "signals" to processes with the form . Notice that sending and receiving those "signals" is greatly simplified by using *tell*, *ask* and the CCP *store*.

*Main*(*edges*,*a*,*b*)
 ()
| **tell**
| **tell**

Following, we give an intuition about how this system works. To find a path between the vertices 1 and 5 in figure 15, it starts by sending *forward* "signals" to all the arcs processes with the form  and *back* "signals" to all the processes with the form . In this example, and receive a *forward* and a *back* "signal" respectively. Concurrently, sends *forward* "signals" to and , to , and to . In the other hand, sends *back* "signals" to and to . As soons as an receives a *back* "signal" and a *forward* "signal", it tells the store that there is path (i.e **tell** () ).

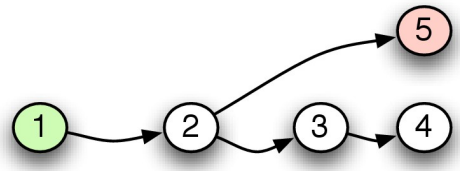

0.8

Figure 15: Example of finding paths in a graph concurrently (1)

Additionally, the reader may notice that there is not a path between vertices 1 and 5 in figure 16. In this example, the *back* "signals" sent to processes are not received by any process. Therefore, none of the receives a *back* and a *forward* signal.

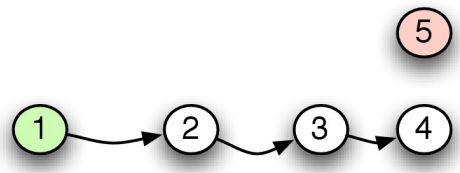

0.8

Figure 16: Example of finding paths in a graph concurrently (2)

After calculating a *fix point*, we can ask the constraint system for the value of . If the variable is not bounded, we can infer that there is not a path.

## 4.2  Implementation using threads

> *The observation to be made (when using threads) is that you always should first propagate to a fixpoint and then perform the ask. While not incorrect (asks are monotonic) it would be inefficient anyway. Christian Schulte*

The first alternative we tried for the CCP interpreter was using *pthreads* (a *medium weight* portable threading library for Mac OS X, Linux and Windows) for the concurrency control and GECODE for the constraint system. In the *pthreads* implementation, all the *tell* and *ask* processes run in different threads and the access to the *store* (represented by variables in a GECODE space) is serialized using locks.

The *assigned* function is used by the *ask* threads to find out when the GECODE Boolean variable, representing their waiting condition, is assigned.

```
bool assigned(BoolVarArgs root, int Pos)
{ bool answer;
  pthread_mutex_lock (&mutexStore); answer = root[Pos].assigned();
pthread_mutex_unlock (&mutexStore);
  return answer;}
```

We have also made an implementation of the graph path problem in common LISP using a CCP interpreter based on *Lispworks* processes [5]. The architecture of both implementations can be observed in figure 17.

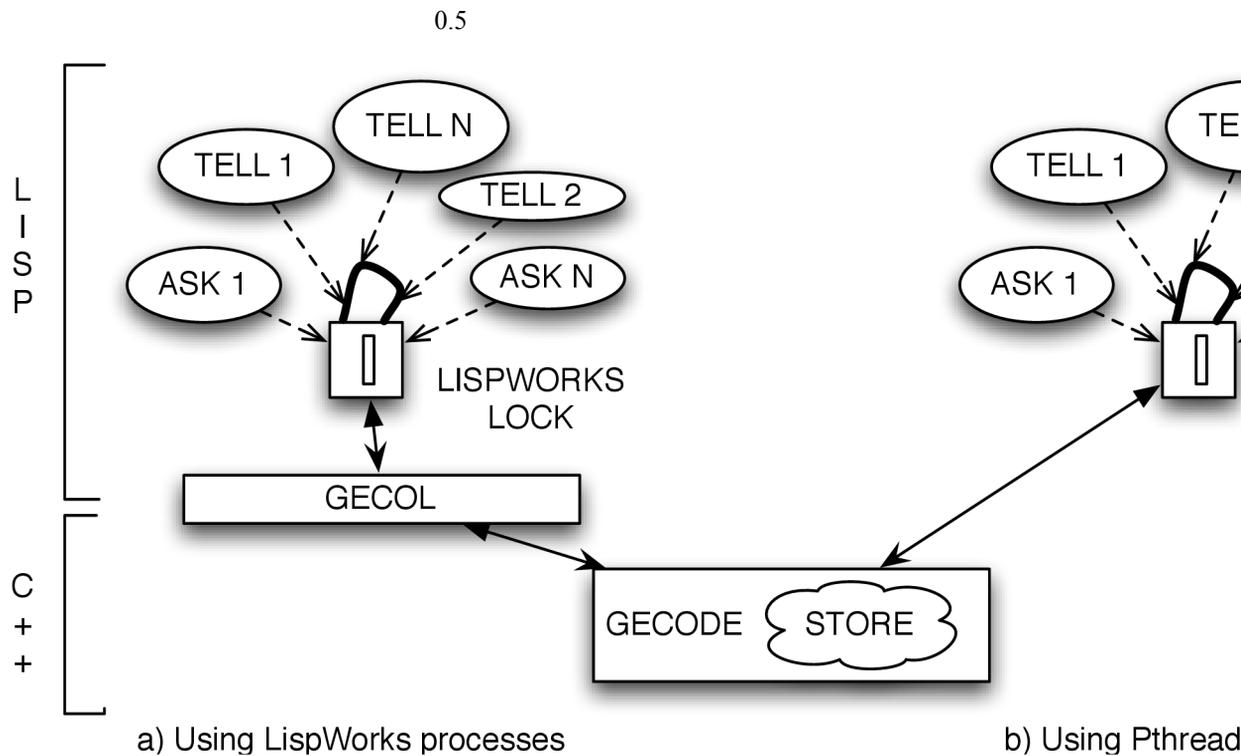

a) Using LispWorks processes          b) Using Pthread

Figure 17: Architecture of the CCP interpreters using Lispworks and using C++

This is the *assigned* function in Common LISP

```
(defun assigned (pos)
 (mp:process-lock thelock) (let ((ret (gecol:boolvar-assigned
(gecol:gecolspace-getbool-int s pos))))
  (mp:process-unlock thelock)  ret))
```

5 LispWorks processes are medium weight threads provided by Lispworks

The *tell* processes are represented directly by GECODE propagators, in this case we use the *rel* propagator to represent the equality constraint (i.e. *tell*(*a*=*b*)).

```
(defun tell (who what)
 (mp:process-lock thelock)
 (gecol:rel-boolvar-intreltype-int-intconlevel
  (gecol:boolvar-val (gecol:gecolspace-getbool-int s who)) :irt-=
what :icl-def)
 (mp:process-unlock thelock))
```

The *FowardWhen* function represents **when do tell** (). The implementation using pthreads is

```
void * FowardWhen(pair * param)
 {int i = param->i; int j = param->j;
  while ( not assigned(root,i) ){Gecode::Space::status (); };
  Gecode::Space::status ();
  if (val(root,i) == 1)

{pthread_mutex_lock(&mutexStore);rel(this,root[j],IRT_EQ,1);pthread_mut
ex_unlock(&mutexStore);}
  pthread_exit(0);}
```

and this is the implementation using LispWorks processes

```
(FowardWhen (lambda (i j)
 (loop while (not  (assigned i)) (gecol:gecolspace-status s))
 (gecol:gecolspace-status s)
 (if (equal (gecol:boolvar-val (gecol:gecolspace-getbool-int s i)) 1)
(tell j 1))))
```

The drawback of both implementations is the inefficiency of using the *status* [6] each time we want to query if the variable is assigned, since GECODE *propagators* are *lazy* (i.e. they act by demand). Making extensive use of the *status* function would be inefficient event if we use an efficient *lightweight threads* library such as *Boost* (http://www.boost.org) for C++ .

## 4.3 Event Driven CCP interpreter in Common Lisp

We chose the *event-driven* model for the implementation of the CCP interpreter, because we do not use synchronous I/O operations and all the synchronization is made by the *ask* processes, using constraint entailment. This implementation still makes extensive use of the *status* function, but we have changed the model of concurrency. Instead of using *pthreads* or *LispWork processes*, we use *event driven programming*. We have 3 types for events of CCP: *ask*, *tell* and *parallel* (see figure 18).

---

6 *status* is a GECODE function used to calculate a *fixpoint*

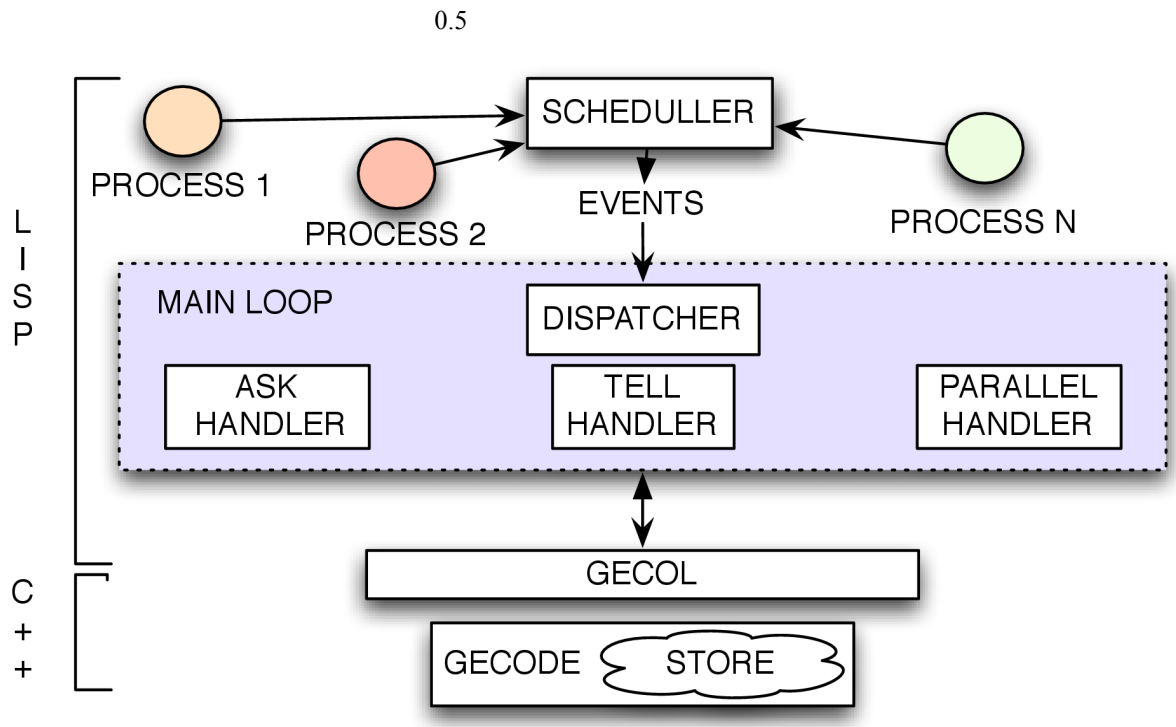

Figure 18: CCP interpreter using Event Driven Programming and Gecol 2.

The handler for the *ask* events checks if the boolean variable, representing their waiting condition, is assigned. If it is not assigned, it adds the same *ask* event to the queue, else it checks the value of the boolean variable. If the boolean variable is true it adds the continuation of the *ask* process to the event queue, else no actions are taken.

```
(defun askHandler (Var Body)
 (if (gecol:boolvar-assigned (gecol:gecolspace-getboolt-int *store*
Var)) ; is assigned?
    (let () (if (== (gecol:gecolspace-getbool-int *store* Var) true)
        (setf (thread-event *current-thread*) Body) ;next event is body
        (setf *current-thread* nil))) ;else there are no more events
    (let ()  ;if the variables are not assigned, create ask event again
        (setf (thread-Event *current-thread*) (make-ask :Vars Vars :Body
Body))))
```

The **when do tell** () process is represented in this interpreter as follows

```
(make-ask :Lvars (list   i)    ; FowardWhen
  :Relation (lambda () (equal (gecol:boolvar-val (gecol:gecolspace-
getbool-int *store* i)) 1))
  :Body (make-tellequal :who j :what 1))
```

Using *event driven programming* led us to a faster and easier implementation of CCP than the approaches presented before. But, we realized that instead of creating handlers for *tell*, *ask* and *parallel* and a *dispatcher* for processing the events, we could improve the interpreter's performance taking advantages of the *dispatcher* and event queues provided by GECODE.

## 4.4 Event Driven CCP interpreter in C++

*Why do you write your own dispatcher and handlers*

*when GECODE has them inside and they are very efficient? Gustavo Gutierrez.*

After considering multiple solutions, this is a generic implementation of the CCP interpreter capable of real-time. The *tell*, *ask* and *parallel* processes are represented by classes.

We defined an *AskBody* class, which is a superclass for the *tell*, *ask* and *parallel* classes. This way we can pass any object inhering from this class to the *ask* propagator, making it generic. We do not use function pointers, because then it would be also required to pass the arguments to those functions and it will be less generic.

```
class AskBody
{ public:
  virtual void Execute(Space * h) { };};
```

We also defined an interface (the superclass *tell*) and three classes inhering from it: tellEqual, reprenting *tell* ($a=b$); tellSetIn, representing *tell* ($a \in B$); and tellGE, representing *tell* ($a>b$) (see figure 19). Other kind of *tell* agents can be easily extended inheriting from the *tell* superclass and declaring an *Execute* method. The *Execute* method is called by an *ask* object when a *tell* is nested in an *ask* or it is called by a *parallel* object when it is nested in a *parallel* object.

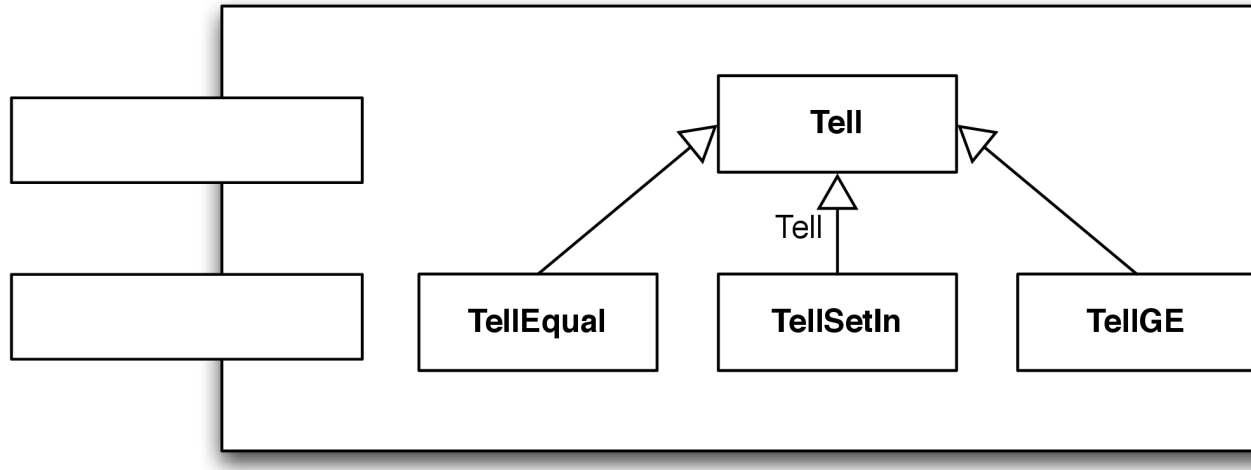

Figure 19: Simplified class diagram of the tell classes

In order to represent the *ask* processes, we have developed a generic *ask* class, with a constructor receiving a pointer to an *AskBody* object and a pointer to a *constraint* object. Both of them are passed to the *ask* propagator, when its *Execute* method is called. The *AskBody* object $P$ is the continuation of the *ask* and the *constraint* object $b$ is the *ask* guard.

These classes inherit from the *constraint* class: SetIn for , EQ for , GQ for , GE for , NOT for , AND for  and OR for  (see figure 20). This can also be extended by inheriting from the *constraint* class and declaring a  method, which returns a GECODE Boolean variable.

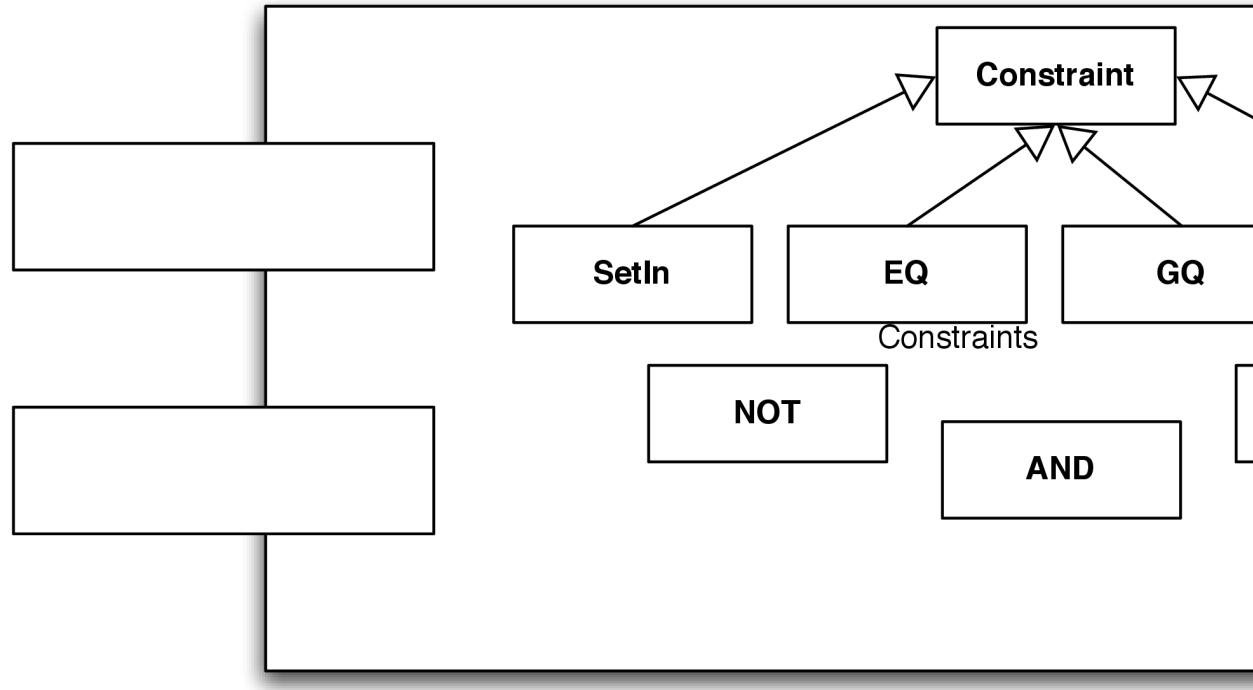

Figure 20: Simplified class diagram for the constraint classes

Once *b* is assigned, the propagator checks its value. For a true value it calls the *Execute* method of *P* (which could be another *ask*, a *tell* or a *parallel*). Then the *ask* propagator will go to the *subsumed* state. Figure 22 show the relation between GECODE, the *AskPropagator* and the *ParallelConditional* (described in next chapter).

```
ExecStatus AskPropagator::propagate(Space* home, ModEventDelta med) {
    if (b.one()) {P->Execute(home); assert(b.assigned()); goto
subsumed;}
    if (b.zero()) {assert(b.assigned()); goto subsumed;    }
    return ES_FIX;
  subsumed:
    return ES_SUBSUMED(this,sizeof(*this)); }
```

Figure 21 compares different interpreters running the program to find, concurrently, a path in graph. We present the execution times of a Common LISP recursive function , an implementation using Concurrent Constraint Programming in Mozart-OZ, an implementation using our own *dispatcher* in Common LISP and the implementation in C++ using the *ask* propagator. The reader can notice that the performance of the interpreter using the *ask* propagator is much faster than all the other ones. Therefore, we recommend using this interpreter for real-time application using the CCP model.

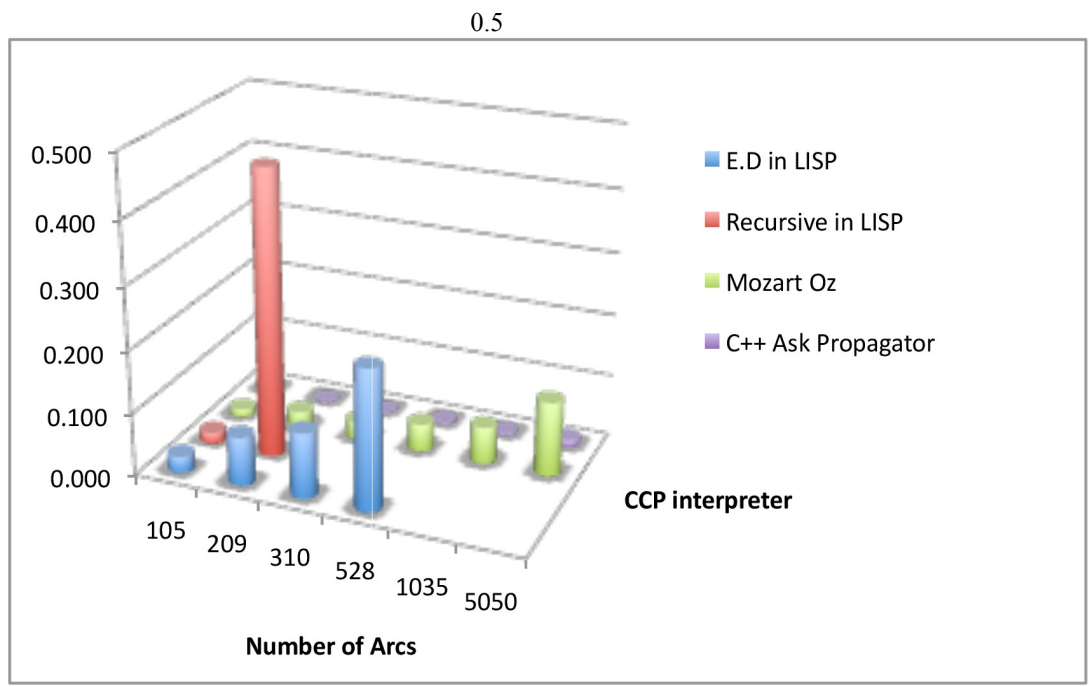

Figure 21: Comparing different CCP interpreters (time in seconds)

## 4.5 Future Work

> *You might not been able to find the optimal path concurrently, but you can find multiple paths, rank them and give a sub-optimal solution with a time limit constraint. Gérard Assayag*

We used the interpreter to run the CCP specification to find a path in graph. It can be easily extended to find as many paths as we can in a certain time, rank them according to a weight provided for each edge, and returning the path with the highest rank. Since we represented the *ask* process as a monotonic propagator, we can use the *BAB* search engine provided by GECODE to control the search and the time objects (e.g. *TimeStop*) to manage the time demands. In the future we will use the CCP interpreter to find musical sequences in the music improvisation system Omax([8]) with a time limit constraint.

## 5 Ntccrt, a a generic real-time NTCC interpreter

> *If you want to achieve real time interaction you should write your interpreters in C++ and then worry about the communication with OpenMusic. Carlos Agón*

The NTCC interpreter extends the design proposed for the CCP interpreter in chapter 4. In order to represent NTCC processes, we wrote a *parallel conditional propagator* to represent the non-deterministic choice and a few new classes.

The *parallel conditional propagator* receives a sequence of tuples , where is a GECODE Boolean variable representing the condition of a *reified propagator* (e.g. ) and (a pointer to an *AskBody* object) is the process to be executed when is assigned to *true*. The propagator executes where . Then, it goes to a *subsumed* state. If all the variables are assigned to false it goes to a *subsumed* state too.

```
ExecStatus ParallelConditional::propagate(Space* home, ModEventDelta med) {
   int falses = 0;
   for(int i=0; i < x.size(); i++)
   { if (b[i].one()) {P[i]->Execute(home); goto subsumed;}
     else if  (b[i].zero()) {assert(b[i].assigned()); falses++; }}
   if (falses == b.size()) { goto subsumed;}
    return ES_NOFIX;
   subsumed:
    return ES_SUBSUMED(this,sizeof(*this));}
```

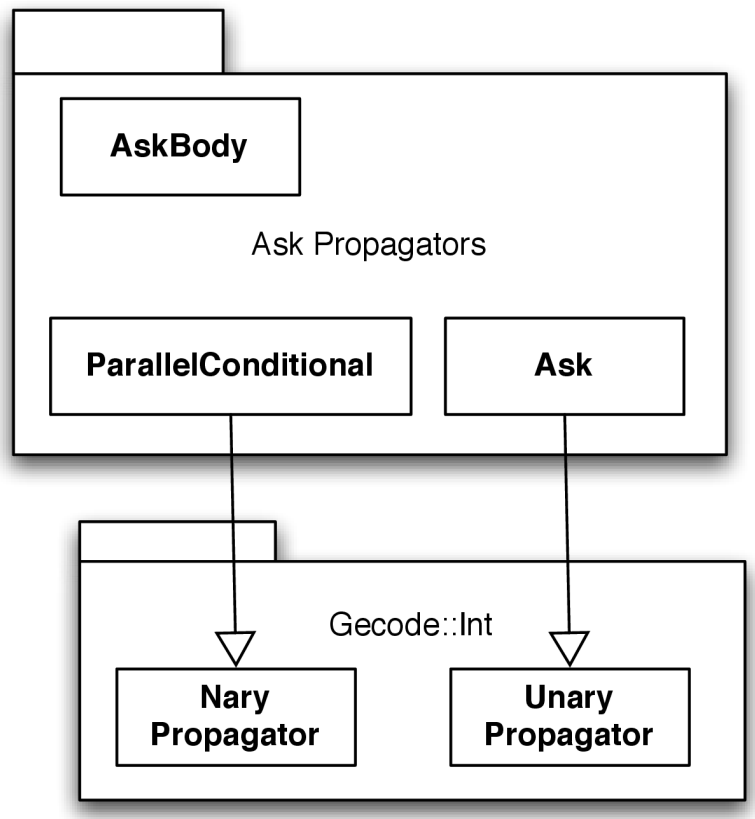

0.5

Figure 22: The Ask Propagator and the Parallel Conditional Propagator

However, the *parallel conditional propagator* has a problem, when there are multiple GECODE Boolean variables assigned *true*, it always chooses the firts one. For that reason, the *non-deterministic* class receives a sequence and calls the *parallel conditional propagator* with a new sequence $R'$ containing the same elements of $R$ in a random order.

## 5.1 Data Structures in the NTCC interpreter

We still use the *tell*, *ask*, *parallel* and *constraint* classes from the CCP interpreter. Additionally, we added some classes to model: the variables, the NTCC *store*, the agent, the procedure calls, the procedure definition, a *skip* class for debugging and the time processes.

Each class inheriting from *TimeProcess* (*NextN*, *Bang*, *Star* and *Unless*) defines an *Execute* method, since *TimeProcess* inherits from *AskBody*. For example, the *Star*'s *Execute* method randomly chooses the time unit to execute the *AskBody* object and place it in the corresponding *process queue*. The *TimeProcess* is an abstract class providing the following information:

- *current time unit* is an integer to know the current NTCC *time unit*.

- *unless queue list* is a pointer to a data structure, where all the *unless* processes wait for their execution at the end of each *time unit*.

- *process queue list* is a pointer to a data structure where all the processes are stored to be executed each *time unit*.

- *continuation process* is a pointer to an *AskBody* object, corresponding to the continuation of the process (e.g. $P$ is the continuation of !$P$).

- *persistent queue list* is a pointer to a data structure, where all the *persistent assignation* processes wait for their execution at the end of each *time unit*.

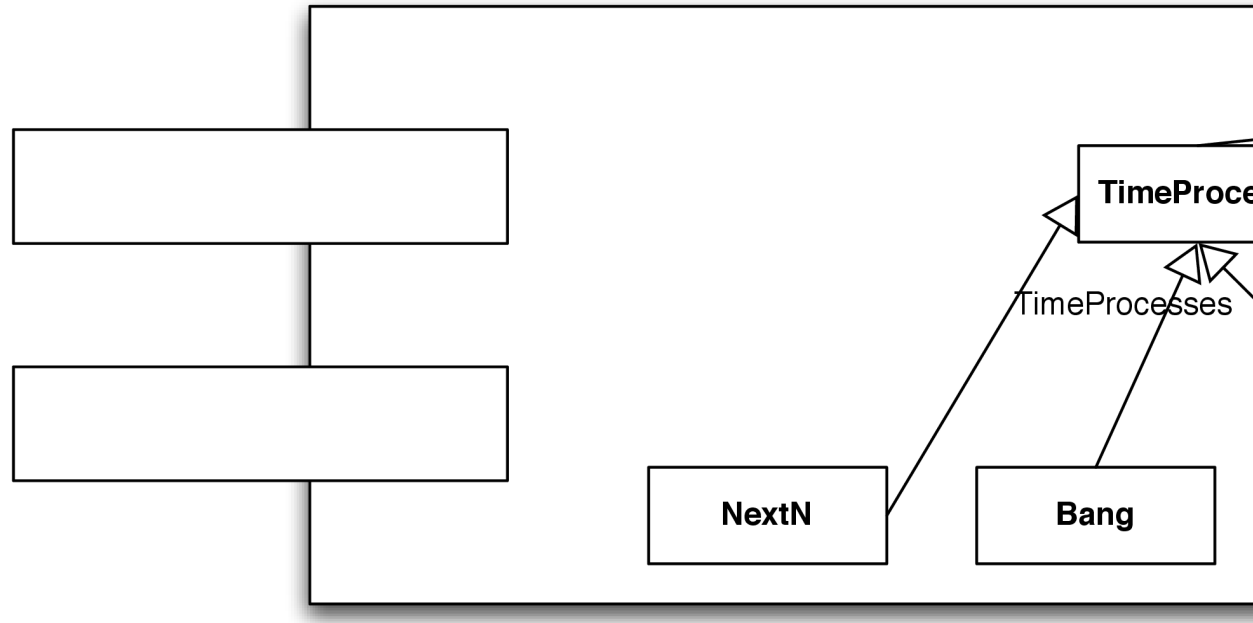

Figure 23: Time processes

In NTCC, we create a *NTCCSpace* (inheriting from GECODE Space) for each *time unit* and the variables are created again for each *time unit* [7]. For that reason, we have our own variable classes and a *store* class in charge of creating the GECODE variables, required for each *time unit*, in the corresponding space (see figure 24). The variable classes are used to model different constraint systems: *BoolV* and *IntV* are used to model the finite domain (FD), *SetV* is used to model finite set domain (FS), and *SetVArray*, *BoolVArray*, *IntVArray* are used to model the rational trees.

---

[7] In the CCP interpreter, since we use GECODE variables directly, we only have one *store* (using a GECODE space to represent it).

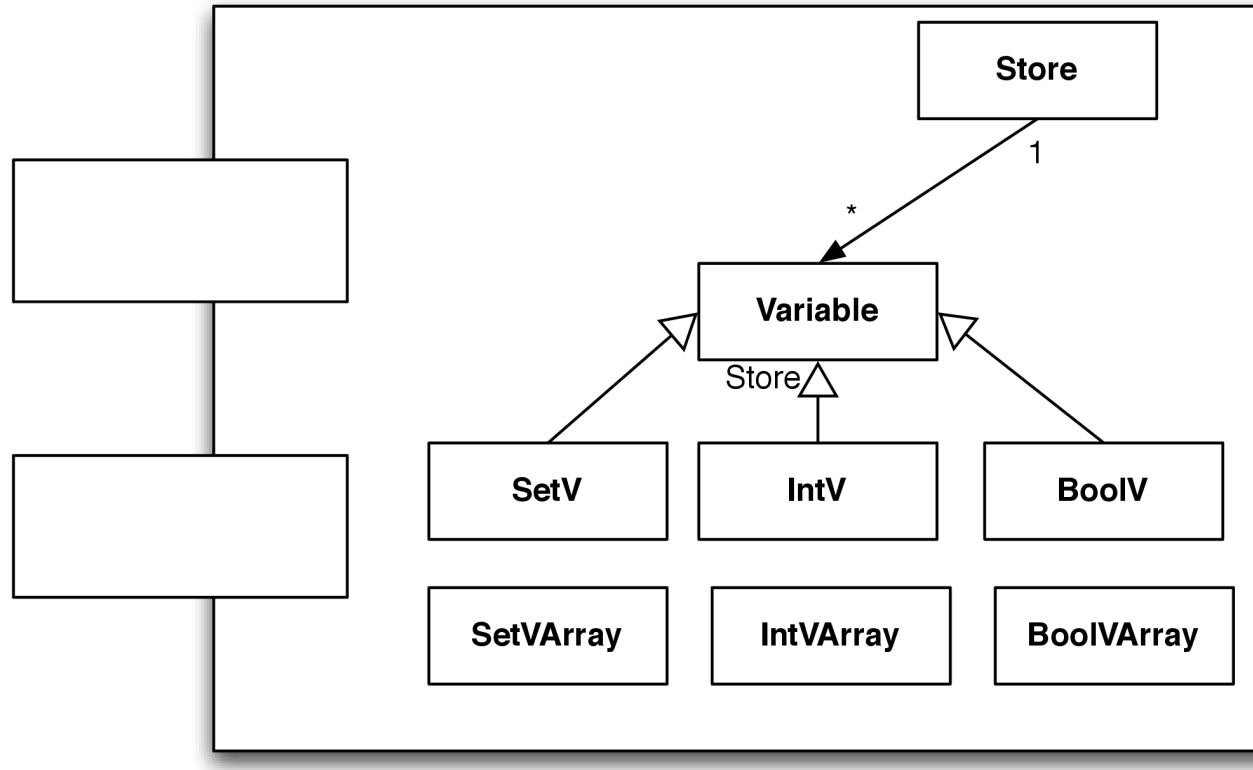

Figure 24: The NTCC Store

We have a *procedure* class used to model both, NTCC simple definitions and NTCC recursive definitions, which are invocated using the *call* class. We also provide some abstractions to represent processes such as !*tell*(*a*=*b*) or !*tell*(*a*∈*B*), called *persistent assignation*. These processes required an special handling, because the variables associated to them are created again for each *time unit*.

In order to execute all these processes, we have three queue lists (proving a queue for each *time unit* to be simulated): *unless* and *persistent assignation* to execute the corresponding processes at the end of each time unit and *process* to execute all the other processes during each time unit. Figure 25 presents the complete data model of the interpreter.

0.5

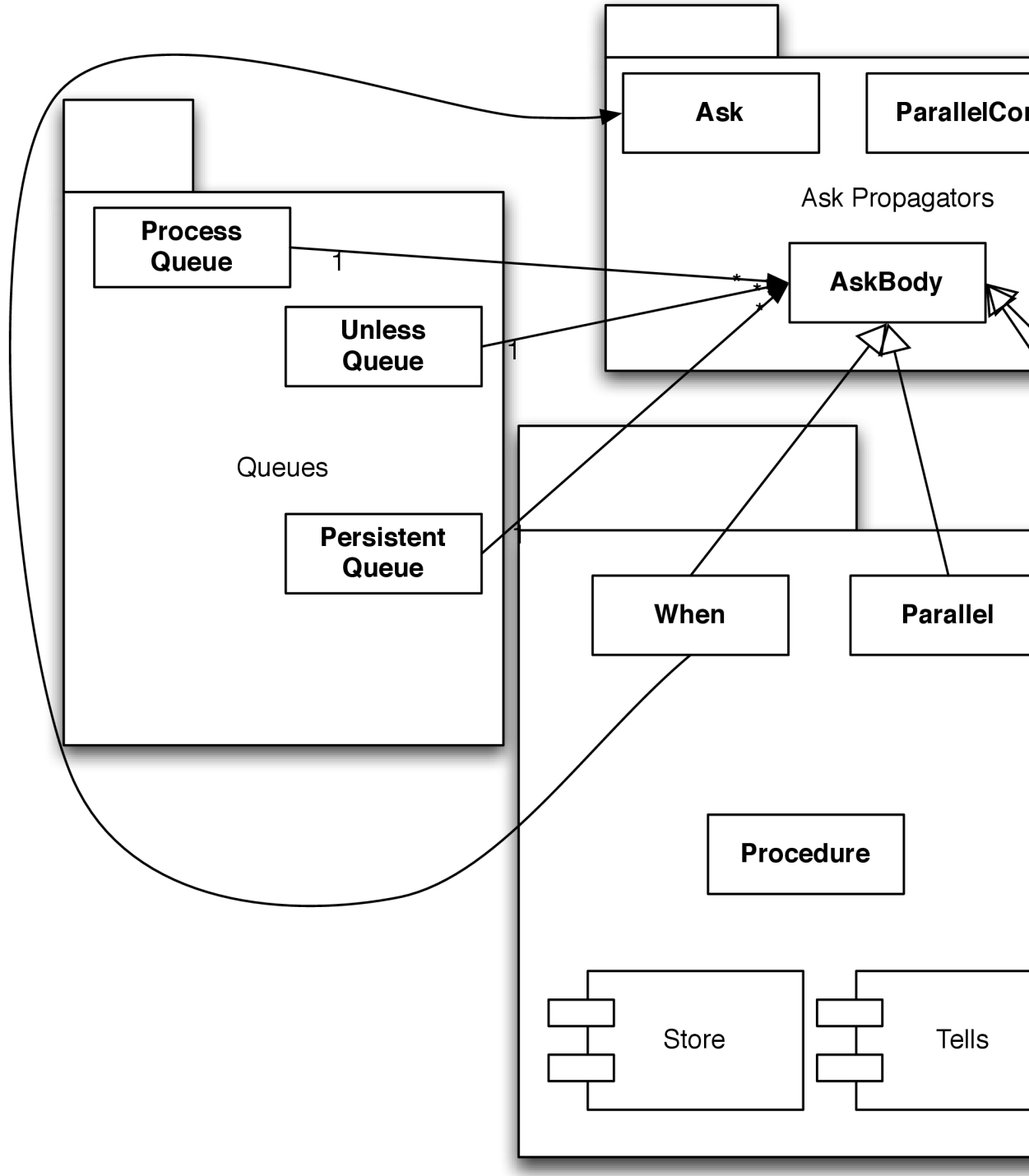

Figure 25: The NTCC classes

## 5.2 Execution model

To run a NTCC specification the user must define two methods from the *NTCCSpace*, one for getting the input from the environment (e.g. a midi keyboard) and the other one for sending an output to the environment (e.g. printing in the screen or sending pitches to OpenMusic). After defining those methods, the user writes the NTCC specification and compiles the program.

In order to execute the simulation, the user calls the compiled program with the number of units to be simulated and the parameters of the main NTCC definition (if any). Then, for each time unit $i$ these instructions are executed (see figure 26):

1. Create a new *NTCCSpace*.

2. Create a new *store* and new variables.

3. Call the *input processing* method.

4. If $i=0$ execute the main NTCC definition with the arguments given by the user.

5. Move the *unless* processes to the *unless queue i*.

6. Move the *persistent assignment* processes to the *persistent assignment queue i*.

7. Execute all the remaining processes in the *process queue i*.

8. Calculate a *fixpoint*.

9. Execute the *unless* processes in the queue $i$.

10. Execute the *persistent assignments* in the queue $i$.

11. Call the *output processing* method.

12. Delete the current space.

0.5

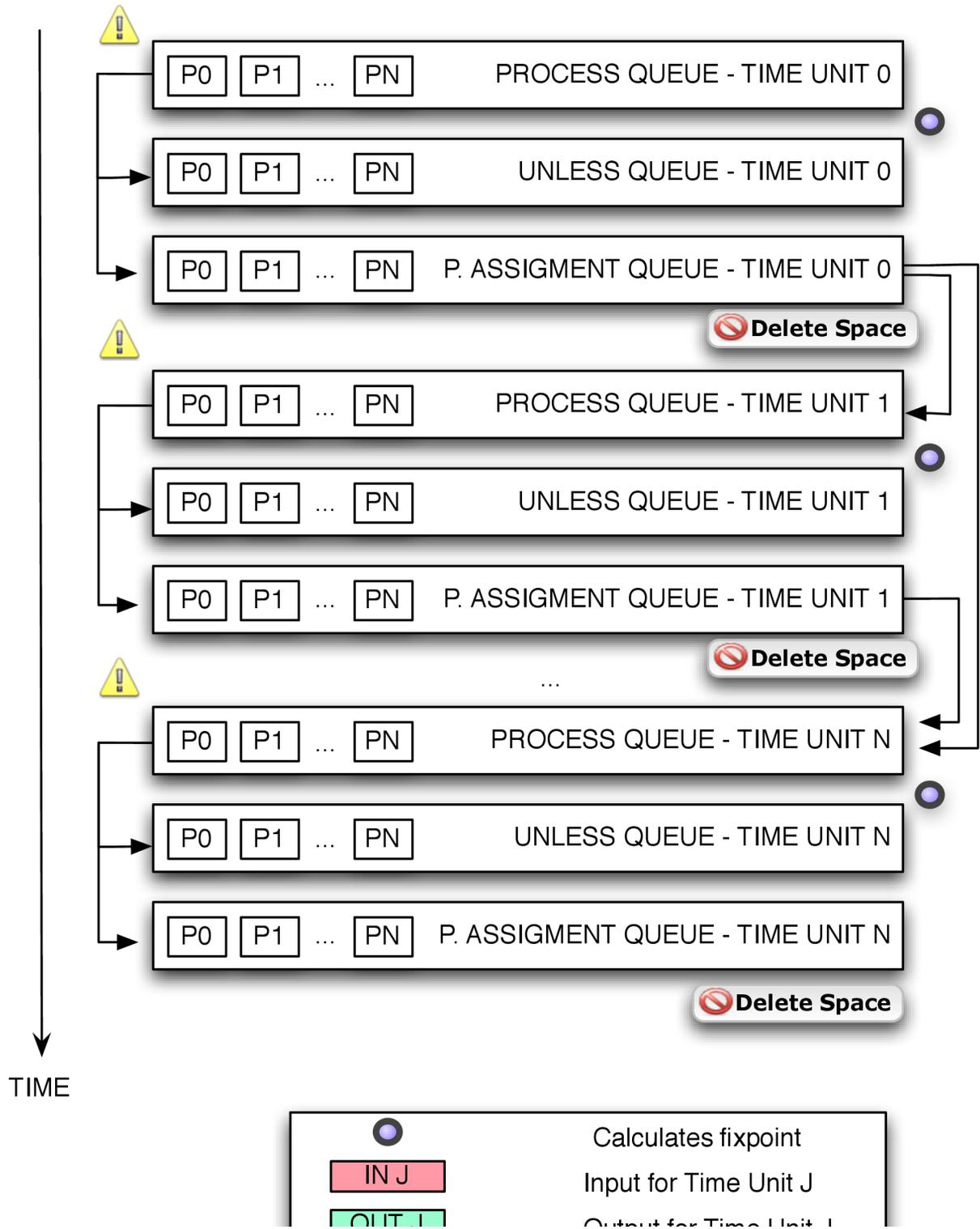

Figure 26: Execution model of the NTCC interpreter

## 5.3 Possibilities and limitations of the interpreter compared with NTCC

The NTCC interpreter offers two features not found in the NTCC formalism. It is able to express general recursion (e.g. it can make multiple recursive calls in a recursive procedure) while the NTCC formalism offers a restricted kind of recursion. Additionally, since we have encoded the *ask* as a GECODE propagator, we are able to use search in some NTCC[8]. Unfortunately, in order to achieve real-time, the interpreter is not able to execute certain processes such as:

- !!(*tell*($A=1$)+*tell*($A=2$)) because they make the simulation inconsistent.

- !*tell* ($a \in C$) or *tell* ($a>b$), because in this case it is necessary to copy the domain the variable $C$ in the next time unit [9].

- *tell* ($a>b$) | !*tell* because we have written a propagator for the rational trees, in charge of waiting until $a$ takes a single value and it does not happen in this process.

- , because we represent those variables with GECODE arrays (using a matrix lineal representation), therefore we need to know at least one dimension of δ.

Another limitation of *Ntccrt* is the restriction over the domains for the different constraint systems. The domain for FD variables is [−2147483645,2147483645]. For FS and the rational trees each data structure can have up to 2147483645 finite domain variables. This limitation is due to GECODE, which uses the C++ integer data type for representing its variables.

## 5.4 Applications: Concurrent Constraint Factor Oracle for Music Improvisation

*The idea of writing the FO in NTCC is not replacing Omax.*

*The idea is showing that constraints and blocking ask provides a synchronization mechanism that would be difficult to write in Omax. Camilo Rueda*

NTCC procedures are written in the interpreter in a declarative and intuitive way. For each procedure in the model (e.g. ) it is necessary to declare and instantiate a class inheriting from *procedure*, where the *Execute* method is overloaded to receive the arguments and return the resulting process. To use the rational trees constraint system we use the  method provided by the *store* class, allowing us to reference an element in the rational tree. For example, the element in the position $i-1$ of the variable *S* can be referenced as *thestore*->. Once the element is referenced, we use it as we would do with a FD variable (*IntV*). Following this intuitive syntax, the

---

8 Models using non-deterministic choices are incompatible with the recomputation used in the search engines

9 Allowing generic *persistent assignment* will be a major change in next release of Ntccrt.

process, in charge of the synchronization between the  and the  processes, is written as

```
Gecode::Int::AskBody * syncp::Execute()(Space * h,vector<int>
intparameters,
 vector<variable *> variableparameters ){
  int i = intparameters[0];
  return ntcc::parallelp(  ntcc::whenp(ntcc::ANDc(ntcc::GQc(thestore-
>create_IntV(S,i-1,h), -1),
                                ntcc::GQc(go,i)),
ntcc::parallelp(ntcc::callp(Add,i),
ntcc::nextnp(ntcc::callp(Sync,i+1)))),
                        ntcc::unlessp(ntcc::ANDc(ntcc::GQc(thestore-
>create_IntV(S,i-1,h), -1),
                                ntcc::GQc(go,i)),
ntcc::nextnp(ntcc::callp(Sync,i))) );}
```

We ran this model in an Intel 2.8 GHz using Mac OS 10.5.2 and GCC 4.1, taking an average of 20 milliseconds per time unit, scheduling around 880 processes per time unit. In the other hand, Rueda's interpreter ran the improvisation model in a 1.67 GHz Apple PowerBook G4 using Digitool's MCL version of Common Lisp, taking an average of 25 milliseconds per time unit, scheduling around 20 concurrent processes. Unfortunately, Rueda's implementation uses some MCL's functions (not defined in the Common Lisp standard) and we have not been able to run its interpreter in Mac OS X Intel to compare them.

## 5.5 Future Work

*If you cannot split your interpreter into functions and call them from OpenMusic, then you can use the OpenSound Control (OSC) protocol to comumnicate the interpreter and OpenMusic. Jean Bresson*

In next release of Ntccrt we will have an efficient representation of cells, based on some ideas from Rueda's interpreter. We are also interested in providing support to a probabilistic extension of NTCC. Currently there are two probabilistic extensions: *SNTCC* [32] developed by Olarte and Rueda, and *PNTCC* [37] developed by Perez and Rueda. We will have a more efficient representation of the rational trees constraint system and a generic way to represent *persistent assignation*. Additionally, we will be able to write NTCC specifications in Common LISP syntax. For example the  process used for modelling FO, could be represented as

```
(defproc Sync (i)    (||(when (and (v>= S[(- i 1)] -1) (v>= go i))
                     (|| ( Add i) (nextnp ( Sync (+ i 1)))))
                      (unless (and (v>= S[(- i 1)] -1) (v>= go i))(
Sync i))))
```

An interface with OpenMusic will also be done in order to visualize the input and output scores, as well as writing the specification of NTCC using a visual language (see in figure 27). Using the *OpenSound Control* (OSC) protocol, we can communicate the input and the output of *Ntccrt* with *OpenMusic*.

0.5

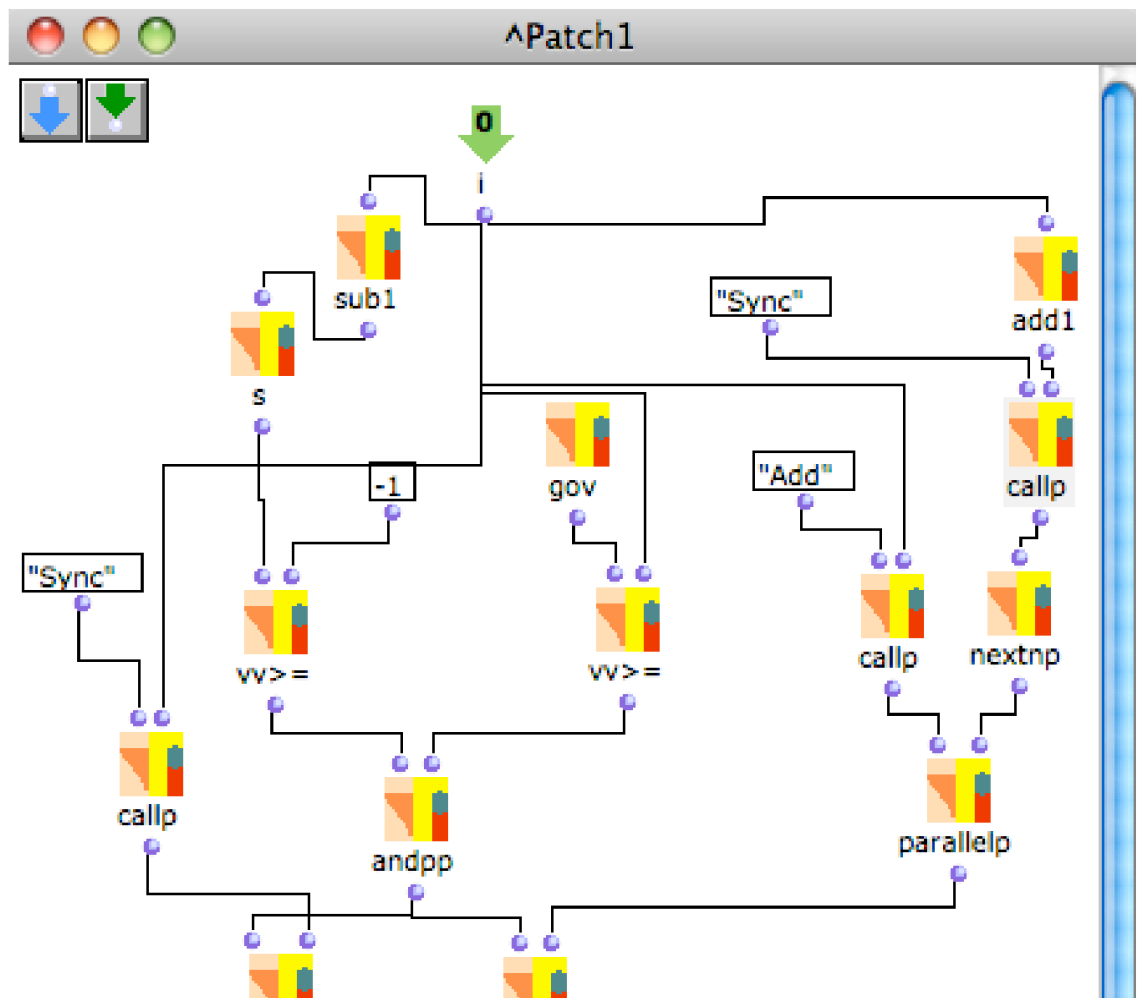

Figure 27: Sync Process in the NTCC interpreter linked with OpenMusic

# 6 Gecol 2 : Gecode 2 Wrapper for Common Lisp

GECOL [52])is a wrapper for GECODE 1.3 versions maintained by Killian Sprotte, providing propagators for finite domain (FD), finite domain sets (FS), the Deep-First-Search (DFS) and Branch-and-Bound (BAB) search engines.

GECOL 2 ([68]), the library we have developed, is an extesion of GECOL maintained by Mauricio Toro Bermúdez, supporting GECODE 2.1.1 (current version of GECODE) and including further support for FS constraints. GECODE 2 is a low level API wrapping the propagators and the search engines mentioned before (see figure 28).

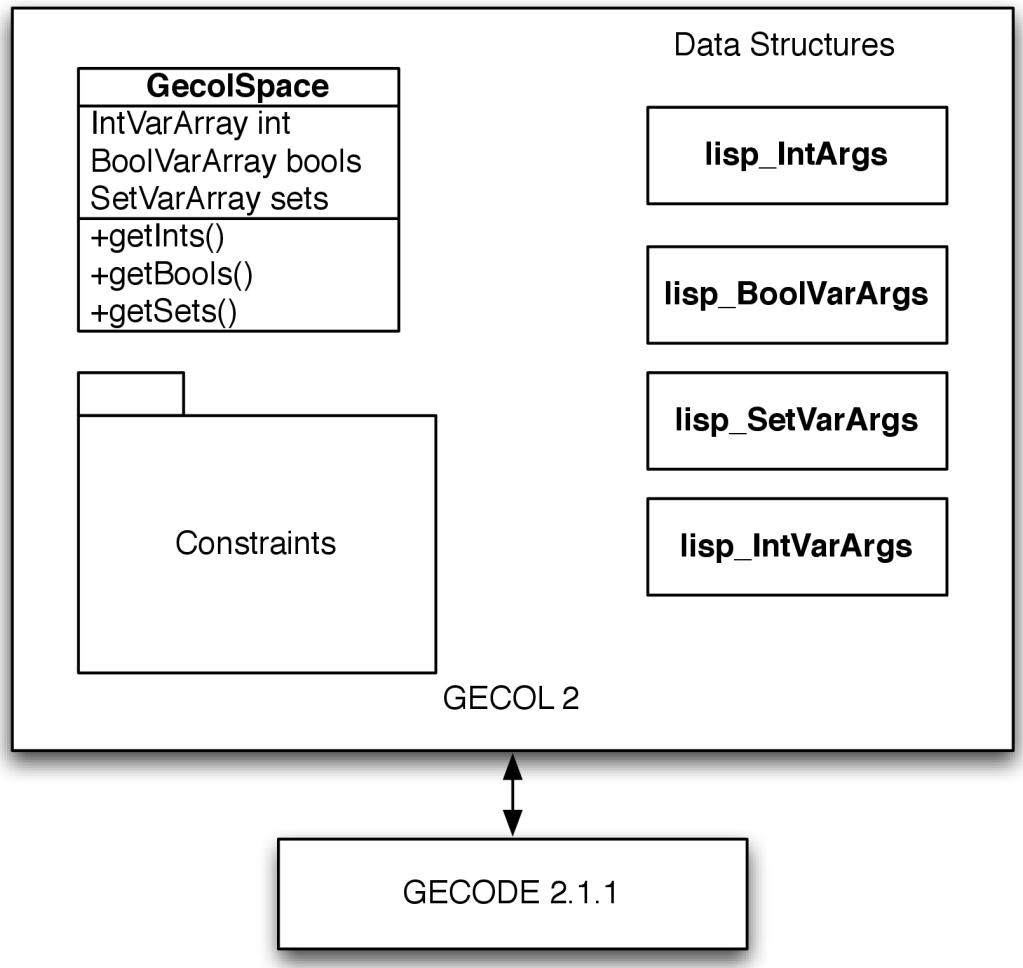

Figure 28: Architecture of the Gecol 2 library

In order to write a finite domain CSP in GECOL 2, it is required to create a *gecolspace* (a class inheriting from GECODE's space), declaring the number of variables to be used and their the domain. Then we add the constraints and specify the *branching*. Finally, we provide the functions *DFS* and *BAB* to get the solutions and a simple loop allow us to search as many solutions as we want. Following, we present a program to find the solutions for *n* numbers, which sum is 14.

```
(defun add-14 (size)
  (let ((s (gecol:make-gecolspace :intnum size :intmin 0 :intmax 11)))
    (gecol:with-var-arg-array ((loop for i below size collect
                   (gecol:gecolspace-getint-int s i)) varargs)
      (gecol:linear-intvarargs-intreltype-int-intconlevel s varargs
:irt-= 14  :icl-def)
```

```
    (gecol:branch-intvarargs-bvarsel-bvalsel s varargs :bvar-none
:bval-min))
    (let ((e (gecol:make-dfs-space-int-int-stop s)))
      (loop for sol = (gecol:dfs-next e) until (cffi:null-pointer-p
sol)
         do  (dotimes (i size) (format t "~a," (gecol:intvar-val
(gecol:gecolspace-getint-int sol i)))))))))
```

We also provide an interface for FS. This example shows how can we write a program that calculates the difference between two sets *A*={1,2,3,4,5} and *B*={3,5,6,7,8}.

```
(defun setdif ()
    (let* ((s (gecol:make-gecolspace :setnum 1) )
    (a (gecol:gec-fs-make s)) (b (gecol:gec-fs-make s))
    (c (gecol:gecolspace-getset-int s 0)) )
      (gecol:dom-setvar-setreltype-int-int s a :srt-eq  1 5)
      (gecol:dom-setvar-setreltype-int-int s b :srt-eq  3 8)
      (gecol:rel-setvar-setoptype-setvar-setreltype-setvar s a :sot-
minus b :srt-eq c )
      (gecol:branch-setvarbranch-setvalbranch s :set-var-none :set-
val-min)
    (let* ((e (gecol:make-dfs-space-int-int-stop s)) (sol (gecol:dfs-
next e)) )
         (print (gecol:gec-fs-value (gecol:gecolspace-getset-int sol
0))))))
```

## 6.1  Gecode 2 Vs Gecol 2

*You should define an API on top of GECOL 2, allowing a user which is familiar with CSP, but not with GECODE, to write programs using GECOL 2. Carlos Agon*

We wrote two benchmark examples provided by GECODE 2 in GECOL 2 , the n-queens and *all-distinct* stress examples. The efficient version of n-queens, using *all distinct* constraints, was tested in both libraries in an Intel 2.8 GHz using Mac OS 10.5.2, GCC 4.1, GECODE 2.1.1 and *Lispworks* 5.02 *professional* . The reader can notice in figure 29 that time consumption of GECOL 2 is only about 50% more when using GECODE 2. In the other hand the memory consumption, presented in figure 30, is the around twice compared with GECODE 2.

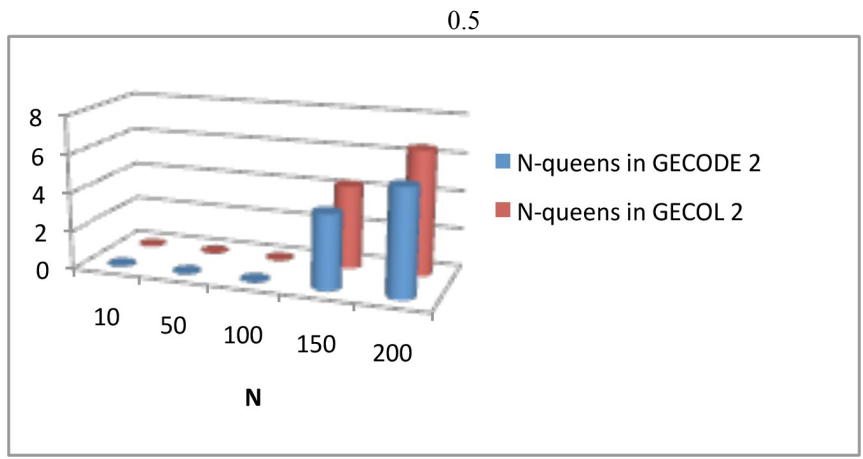

Figure 29: Comparing Nqueens in GECODE and GECOL 2 (time in seconds)

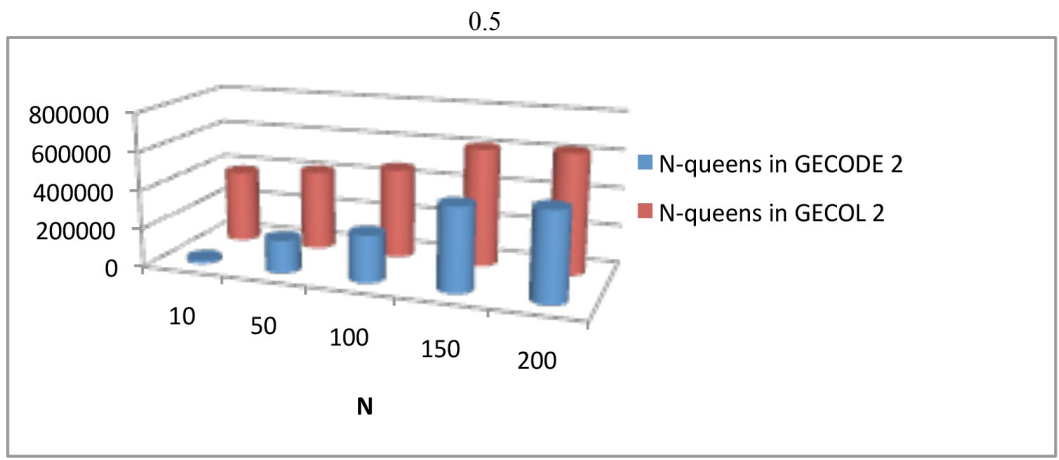

Figure 30: Comparing Nqueens in GECODE and GECOL 2 (memory in bytes)

We also tested the *all-distinct* stress example provided by GECODE 2. As well as the n-queens example, the execution times obtained with GECOL 2 (see figure 31) are only about 50% more than using GECODE 2. Futhermore, constrasting to the n-queens example, GECOL 2 uses an average of 10 times more memory to solve this problem than GECODE 2. The reason is that integer arrays used in GECODE 2 are represented using LISP lists in GECOL 2, leading to high memory consuming when using big arrays (e.g. 50000 element arrays) [10].

---

10 We expect to provide a more efficient representation for GECODE arrays in GECOL 2 in next release.

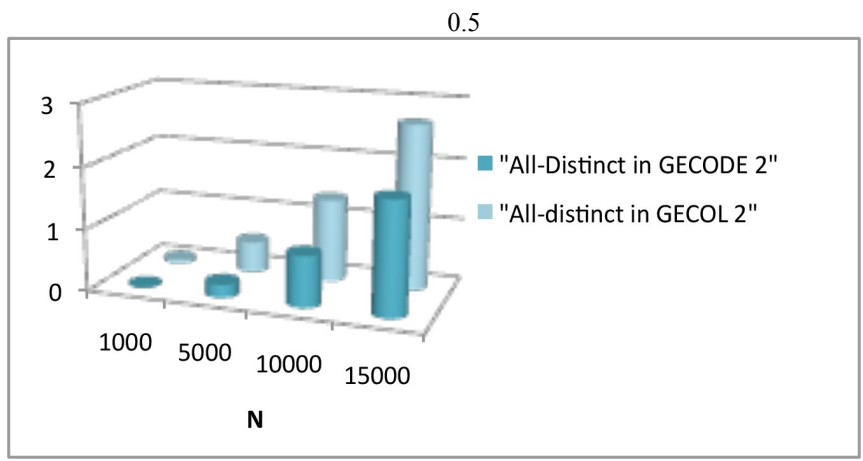

Figure 31: Comparing All-distinct in GECODE and GECOL 2 (time in seconds)

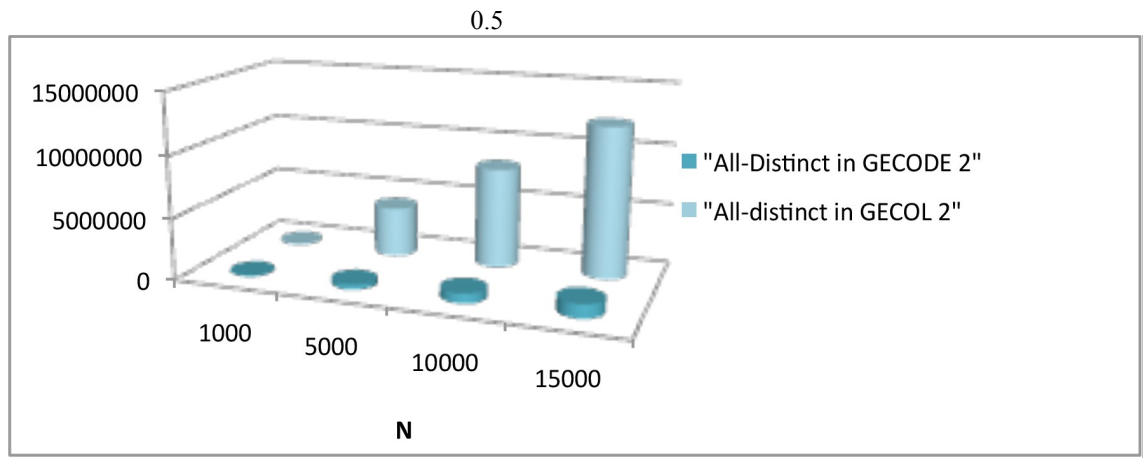

Figure 32: Comparing All-Distinct in GECODE and GECOL 2 (memory in bytes)

## 6.2 Applications: Modelling k-nets

*Traditionally the music analysis using k-nets has been done by hand, Constraint Programming could be a better approach to solve this problem. Moreno Andreatta*

We represented a *k-net* in GECODE and GECOL 2 as an adjacency matrix (a common representation for graphs). Following this representation, we wrote a CSP to find all the solutions. First, we wrote a program in C++ using GECODE and then, in Common LISP using GECOL 2 and *Lispworks* CAPI library (for drawing graphs). GECODE 2 runs around 3 times faster GECOL 2 for solving this problem, when we print the solutions. In the other hand, if we use LISP lists to store all the solutions, time consumption and memory consumption gets very high using GECOL 2.

Formally a CSP is a tuple $<X,D,C>$ where $X$ is the set of variables, $D$ is a domain of values and $C$ is the set of constraints. The input of this problem is a class pitch $I$ represented as a tuple  and $K$ the desired inversions. The variables for the CSP are $X =$ , their domains are $D = \{0,1,2\}$. For the domain we represent when there is not an edge as 0, transpositions as 1 and inversions as 2.

For the constraints, we consider that if there is a transposition or inversion from $i$ to $j$ there is also one from $j$ to $i$, that way we can represent multiple solutions in a single adjacency matrix. The constraints $C$ are the following relations over all the variables in $X$:

- the number of variables distinct from 0 are greater or equal than $2*n$

- the number of variables equal to 2 are $2*K$

- for each $i \in [0..n], j \in [0..n]$ if $i=j$ then

- for each $i \in [1..n], j \in [1..n]$ if $i<>j$ then .

The "for each" constraints can be easily represented in GECOL 2 as follows

```
(dotimes (i n) (dotimes (j n)
    (if (equal i j)
      (gecol:rel... s (gecol:gecolspace-getint-int s (+ (* i n) j))
:irt-= 0 :icl-def)
      (gecol:rel... s (gecol:gecolspace-getint-int s (+ (* i n) j))
                   :irt-= (gecol:gecolspace-getint-int s (+ i (* j
n))) :icl-def))))
```

This application finds all possible *k-nets* for an array of *n* pitches and *K* desired inversions. The sources can be found in the examples provided by GECOL 2. This program will be a part of the application developed for Yun-Khan Ahn's doctoral thesis. The reader can find some results in the slides [2] presented in mamuX 2008. Figure 33 shows solutions for the pitch class <3,10,11> (representing <*Eb,Bb,B*>) and *K*=1. For instance, the last solution , represents in a compact way this solutions:

- 

- 

- 

- 

0.3

```
ergraph-op
ergraph-op

    ))

   :be
   :be

  ))
HYPER

CL-USER 3 >
(hypergraph-openmusic *notes* 1)
((("--" "I1" "T4") ("I1" "--" "T11") ("T8" "T1"
 "--" "I9") ("T8" "I9" "--")))

CL-USER 4 >
```

Figure 33: Solutions for k-nets using GECOL 2

## 6.3 Future Work

*The next step is to find the isographies between graphs, building an hypernetwork.*
*Yun-Khan Ahn.*

GECOL 2 is a low level API, which requires deep knowledge of GECODE in order to use it. GECOL 2 will be the base on which we will integrate the CCP and NTCC interpreters with Common LISP and the music environment OpenMusic [1].We want to integrate GECOL 2 with OpenMusic, defining a high level API for users who know about Constraint Programming but not necessary about GECODE, to program all kind of application in the visual language provided by OpenMusic. For instance, it will be possible writing a program to find numbers which is sum is 14 in a visual way as we can see in figure 34 or using an intuitive syntax

```
(defun add-14 (size)
  (propagate
        ((varargs (gecol:make-array  size  0  11))) (gecol:linear=
varargs 14) (gecol:branch varargs))
  (for each solution do
       (dotimes (i size) (print-array varargs))))
```

0.5

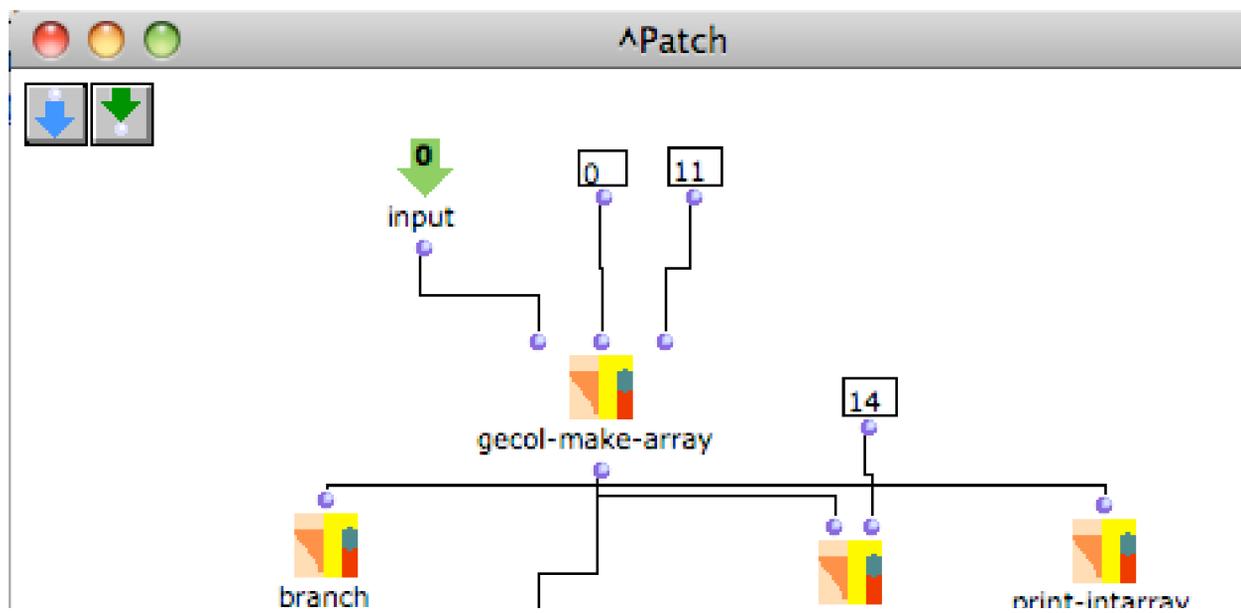

Figure 34: Solutions for a CSP in OpenMusic using Gecol (future work)

The solutions provided for the *k-nets* for 4 notes and 4 inversions are around 10000 solutions. Using the positive and negative isographies explained by David Lewin ([26]), it is possible to build hypernetworks. We have conjectured that using the ask propagator (explained in chapter 4), it will be easy to write a program which computes efficiently all the solutions for the hypernetworks, given a segmented score. This tool will be useful for musicologists and computer scientists doing transformational music analysis.

# 7 Conclusions

- We found out that the time units proposed in *Ntccrt* do not represent discrete time units, because in the simulation they have different durations. This is a problem when synchronizing an NTCC system to other systems. To fix it, we can make the duration of each time unit take a fixed time, allowing us to reason about NTCC time units, as we will do with discrete time units. In order to achieve this, it would be necessary to estimate the maximum duration a *time unit* might last (see figure 35).

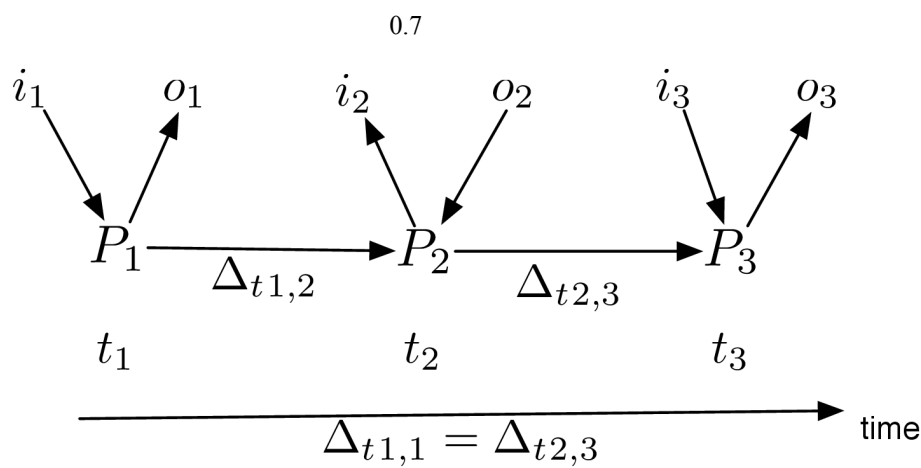

Figure 35: Reactive system in Ntccrt (future work)

- Although GECODE was design to be a library for solving combinatory problems using constraints, we found out that using GECODE for CCP give us outstanding results for real-time. In the other hand, it is very expressive, since most of the propagators used in real-time, have a reified version and those who does not have one, are easily extensible.

- In the future, in order to use *lightweight threads* in Common LISP, we recommend using an implementation with *lightweight threads* such as CMU-CL (http://www.cons.org/cmucl/). Notice that current version of CMU-CL (CMU-CL 19e) provides binaries for Mac OS X PPC and Intel.

- The performance of the interpreter using the ask propagator is much faster than all the other ones. Therefore, we recommend using this interpreter for real-time application using the CCP model.

- We have conjectured that using the ask propagator (explained in chapter 4), it will be easy to write a program which computes efficiently all the solutions for the hypernetworks, given a segmented score. This tool will be useful for musicologists and computer scientists doing transformational music analysis.